\documentclass[%
 reprint,
 amsmath,amssymb,
 aps,
]{revtex4-2}

\usepackage{graphicx}
\usepackage{dcolumn}
\usepackage{bm}
\usepackage{physics}
\usepackage{amsthm}
\usepackage{mathtools}
\usepackage{glossaries}
\usepackage[caption=false]{subfig}
\usepackage{xcolor}
\usepackage{comment}
\usepackage[ruled,vlined]{algorithm2e}
\usepackage[colorlinks=true, allcolors=blue]{hyperref}

\makeatletter
\newcommand*\bigcdot{\mathpalette\bigcdot@{.5}}
\newcommand*\bigcdot@[2]{\mathbin{\vcenter{\hbox{\scalebox{#2}{$\m@th#1\bullet$}}}}}
\makeatother

\makeglossaries

\newacronym{gbs}{GBS}{Gaussian boson sampling}
\newacronym{mzi}{MZI}{Mach-Zehnder interferometer}
\newacronym{svd}{SVD}{singular-value decomposition}
\newacronym{cv}{CV}{continuous variables}
\newacronym{au}{a.u.}{arbitrary units}
\newacronym{mse}{MSE}{mean squared error}

\begin{document}


\title{Realistic simulation of universal multiport interferometers with losses within a Gaussian formalism}

\author{Sebastiano Corli$^{1}$, Dario Chemoli$^{2}$, Samuele Altilia$^{1}$, Edoardo Suerra$^{1}$, Dario Tamascelli$^{1}$, Simone Cialdi$^{1}$, Enrico Prati$^{1,3}$}

\email{enrico.prati@unimi.it}

\affiliation{$^1$Department of Physics Aldo Pontremoli, Università degli Studi di Milano, Italy}
\affiliation{$^2$Department of Physics, Politecnico di Milano, Italy}
\affiliation{$^3$Istituto di Fotonica e Nanotecnologie, Consiglio Nazionale delle Ricerche, Italy}

\date{\today}

\begin{abstract}
Photonic technologies are among the leading platform contenders for demonstrating quantum advantage. Nevertheless, photon losses represent the major hurdle for the realization of a fault-tolerant optical quantum computer, and the need for an accurate characterization of this phenomenon is crucial for classical simulations of quantum algorithms on noisy backends. 
The noisy regime would require the simulation of the  algorithm on a Fock backend or rather assuming the \textit{a priori} hypothesis on the uniformity of the optical losses, which may lead to wrong predictions about the outcome from the algorithm on a cerain hardware.
We propose a Gaussian model that can reproduce a lossy dynamics through the experimental characterization of imperfect interferometers, yielding the corresponding mean vector and the covariance matrix.
Such model can be adapted to any continuous-variable architecture (spatial or time-bin encoded) and for both the Gaussian boson sampling or general-dyne based algorithms.
Eventually, we quantify the drift of optical losses from being uniform throughout the multiport interferometer by a numerical simulation. 
This work proposes a blueprint for future characterizations, tailor-suited for specific devices and architectures, as well as for potential error correction and mitigation techniques, and it paves the way for realistic simulations of quantum algorithms on NISQ optical platforms.
\end{abstract}

\maketitle


\section{Introduction}

%
In the field of quantum computing, photonics is envisaged among the major competitors in the technological landscape. Due to the wide range of degrees of freedom of the photons, many proposals of quantum optical computers have been advanced, e.g. linear optical quantum computers~\cite{knill2001scheme,taballione2021universal,taballione202320,li2015resource,teoh2023dual,li2021heralded}, angular momentum-based devices~\cite{tamburini2026graded} and \gls{cv} devices~\cite{lloyd1999quantum,panichi2026quantum,corli2025quantum,hillmann2020universal,braunstein2005quantum,markidis2022physics,corli2026physical}. However, all these platforms suffer from the same curse, i.e. optical losses of photons and subsequent noise.
%
%
Differently from other technologies, where error correction codes have been designed to correct the flip of the qubit state~\cite{devitt2013quantum,ye2025quantum}, in the frame of quantum optics, which relies on second quantization, the main hindrance consists of the reduction in the population of photons.
An attempt to develop a surface code for \gls{cv} computing has been addressed by Barnes~\cite{barnes2004stabilizer}. However, this framework does not address the main issue in photonic computing, i.e. optical losses. A mitigation technique has been proposed by Su et al.~\cite{su2020error}, leveraging multiple sampling with tunable loss rates in order to reconstruct the ideal target observable to be measured. This mitigation scheme, as aforementioned, heavily relies on the characterization of errors and losses in the optical hardware, and would benefit from an integration with a proper backend for numerical simulation.
Hitherto, differently from the standard error correction codes for qubit-based devices, no technique for error correction against photon losses has been proposed. Quantum compiling is a complementary approach, aiming to reduce the number of ancillary sources as well as the length of the circuit~\cite{kaldenbach2025mapping,corli2026measurement,dang2025qukkos,gupta2026realization,corli2025gauge}.
Nevertheless, much effort has been addressed in characterizing noisy channels and optical losses~\cite{ferraro2005gaussian} due to fabrication defects through the circuitry devices, such as interferometers and filters~\cite{vischi2024simulating}. Even so, up to now, and to the author's knowledge, there is no software or backend for the simulation of \gls{cv} quantum algorithms in presence of noise, differently from libraries suited for matter-based qubits, such as Qiskit or \href{https://quantumai.google/cirq/noise}{Cirq}~\cite{di2023noisy,georgopoulos2021modeling,muqeet2024mitigating,muqeet2025tool,li2022pulse,bravo2024methodology,cignoni2021towards,li2023qasmbench,hu2025suppressing,vorathammathorn2025understanding,biswas2025quantum,misra2026design,pontolillo2025ideal,cattaneo2023quantum}. Moreover, a further distinction has to be underlined between linear optics with single photons~\cite{vischi2024simulating, oszmaniec2018classical} and continuous variables.
The characterization of the optical losses strictly depends on the architecture of the hardware. In the \gls{cv} frame, we may point out two approaches, the spatial-bin encoding and the time-bin encoding~\cite{bacchi2026postselection}. The first one consists of making the light flow through a set of $M$ waveguides, corresponding to the modes of the system~\cite{ranjan2023experimental}. The time-bin encoding approach, instead, relies on a loop-based architecture~\cite{madsen2022quantum,aghaee2025scaling,motes2014scalable,takeda2017universal}, where the wave packets travel in two loops connected by a \gls{mzi} through which the single modes interact with each other. However, the unitary operator to be encoded has first to be decomposed into smaller $SU(2)$ operators, suited for an implementation on a \glspl{mzi} mesh or its time-domain equivalent. The two most adopted schemes are the Clements one~\cite{Clements,hamerly2022accurate}, fitting the spatial approach, and the Reck one~\cite{reck1994experimental}, which has been shown to be equivalent to the loop-based architecture. Other decompositions have been proposed by De Guise~\cite{de2018simple}, where the number of optical elements is diminished from $n \times n$ to $(n-1) \times (n-1)$, and by Liu~\cite{liu2022reduce}, where a 3D architecture is proposed to reduce the footprint of the chip.
A detailed work about the characterization of lossy interferometers in a \gls{cv} loop architecture is provided by Qi et al.~\cite{qi2018linear}. However, this framework can be adapted to any architecture, from time-bin encoded algorithms to spatial meshes.
Photon losses can be represented as the interaction between the external environment and the physical modes from the circuits. Such an interaction is described through a beam-splitter, which in the ideal case -- no losses -- keeps full transmissivity, while in the worst case scenario -- total dissipation -- shifts all the optical information in the external modes. In the most recurrent situation, the global amount of photon losses yields a mixed state, lower losses involving purer states. When introducing external channels to describe each lossy interaction, the number of modes to handle becomes polynomially high, which turns to be computationally demanding when handling Gaussian variables (in the \gls{cv} frame) and mostly unfeasible when directly simulating the Fock space.
Many techniques in literature~\cite{qi2018linear,garcia2019simulating} rely on the \gls{svd} to study the global effect of optical losses on the photonic circuit. This well-known technique consists of decomposing a matrix into a product of two unitaries and, in between, a diagonal matrix displaying the singular values as diagonal elements. Since optical losses cripple the property of the optical transformation to be unitary, the \gls{svd} provides an equivalent representation of the circuit, with lossy channels in the mid dividing two unitary ideal operations~\cite{tischler2018quantum}. Other potentially useful representations are listed and detailed in Ref.~\cite{houde2024matrix}.

%
%
In this work, we address the characterization of optical losses for \gls{cv} quantum computing in a Gaussian formalism. This approach grants us a frame to simulate on a classical backend both experiments of Gaussian boson sampling and algorithms based on homodyne, heterodyne and general-dyne detections.
%
%
We start from the characterization of the imperfect interferometer described by Qi~\cite{qi2018linear} in order to provide a faithful representation of the covariance matrix in presence of the losses characterized for the specific device we are considering. This formalism is adapted to fully describe the \gls{gbs}, where squeezed states are injected into a mesh of \glspl{mzi}, but many other quantum gates in the mid of the computation are engineered to be implemented through an external source injected in the circuit into a beam-splitter, e.g. the displacement gate~\cite{paris1996displacement} or the squeezing in a loop-based architecture~\cite{takeda2017universal,filip2005measurement}.
By leveraging the \gls{svd} the number of external channels involved is minimized, therefore, acting via a partial trace, the covariance matrix of the system is returned. Thanks to such an approach, it is possible to reduce the number of total modes involved (both physical and auxiliary) to $2M$, $M$ being the physical ones in the circuit, instead of a polynomial increase with respect to the length of the algorithm. 
In this work, we describe the full pipeline to simulate the outcome of the quantum algorithm, starting from the characterization of the optical losses on hardware to the final form of the covariance matrix and the probability distribution to be sampled.
An approximation capable of fixing this issue is to consider the optical losses uniform in all channels once the \gls{svd} has been performed~\cite{nehra2026all,oh2024classical,garcia2019simulating}. 
%
%
In the frame we describe, disregarding the depth of the circuit, the \gls{svd} condenses all the lossy channels in $M$ external modes, yielding a covariance matrix of the mixed state with size $2M \times 2M$. The partial trace over this matrix returns the correct statistical moments of the system with minimal effort. 
We showcase some numerical experiments to assess the correctness of such an assumption and, at the same time, the efficiency of our backend.
Furthermore, we leverage this same backend to numerically assess the correctness to assume uniform losses in both Clements and Reck decompositions.

In Section \ref{sec:frame}, we first summarize the theoretical tools to characterize the optical losses in a mesh of \glspl{mzi}, in order to describe the global (non-)unitary operator acting over the modes of the system.
In Section \ref{sec:cov_mat}, we prove how to describe the covariance matrix of an optical system subject to losses, dealing with some specific cases.
In Section \ref{sec:numerical}, some numerical examples are introduced, both to check the correctness of the formalism and to assess whether the optical losses can be approximated as uniform throughout the channels of the multiport interferometer.

\section{\label{sec:frame}Statistics and dynamics in presence of losses}

In this Section, we summarize the theoretical frame to properly characterize optical losses. In the first place, we recall some basic notions of linear optics, introducing the transformations on the first and second moments of the system. Afterwards, we provide the representation of an imperfect interferometer affected by optical losses and the description of CP-maps. Eventually, we sketch a global frame to describe the non-unitary operator (i.e. the operator implemented by a lossy photonic circuit) and some practical examples related to the \gls{gbs}.
\begin{figure*}
    \centering
    \subfloat[]{\includegraphics[width=0.4\linewidth]{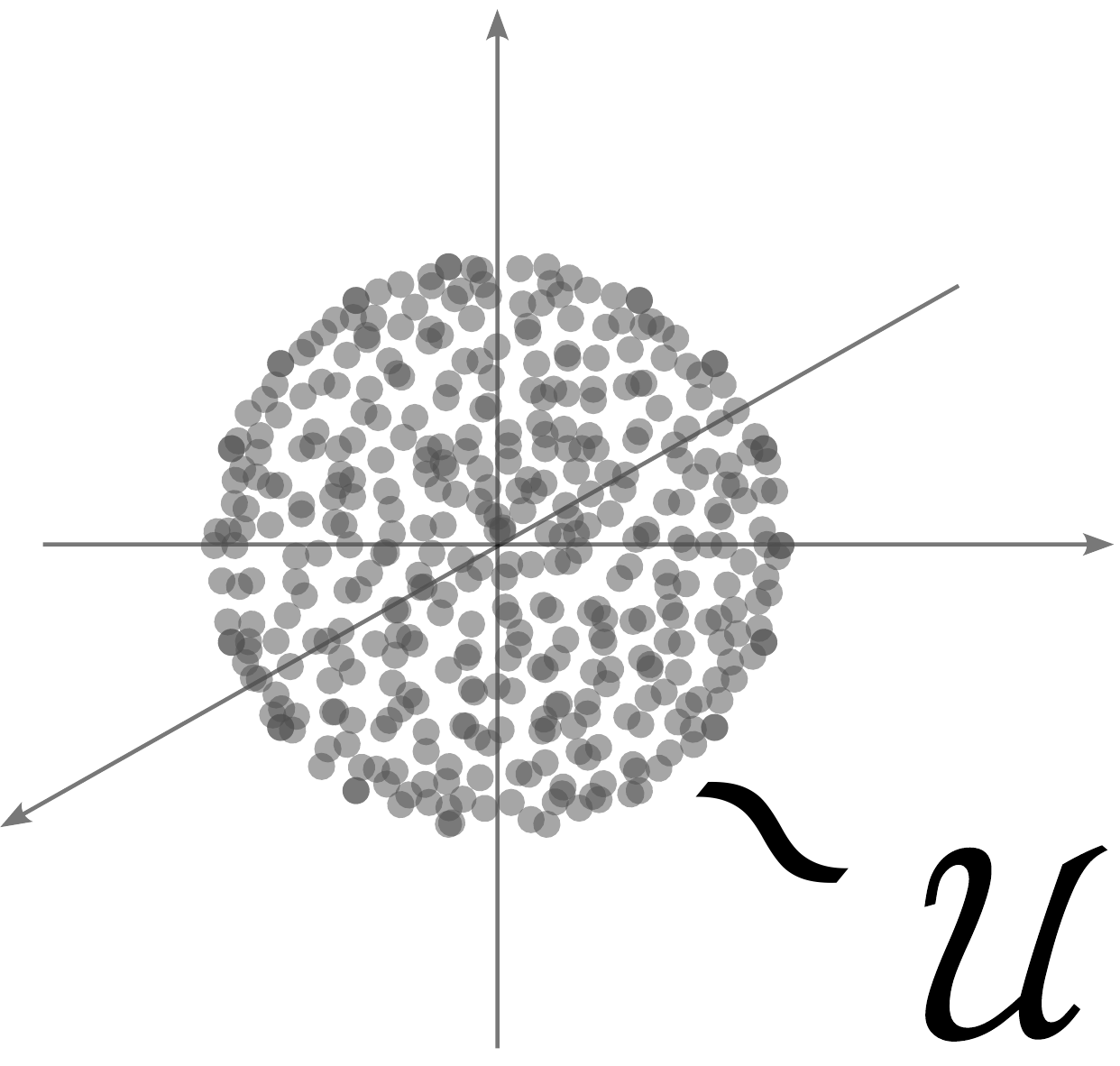}\label{subfig:Haar}}
    \hfil
    \subfloat[]
    {\includegraphics[width=0.35\textwidth]{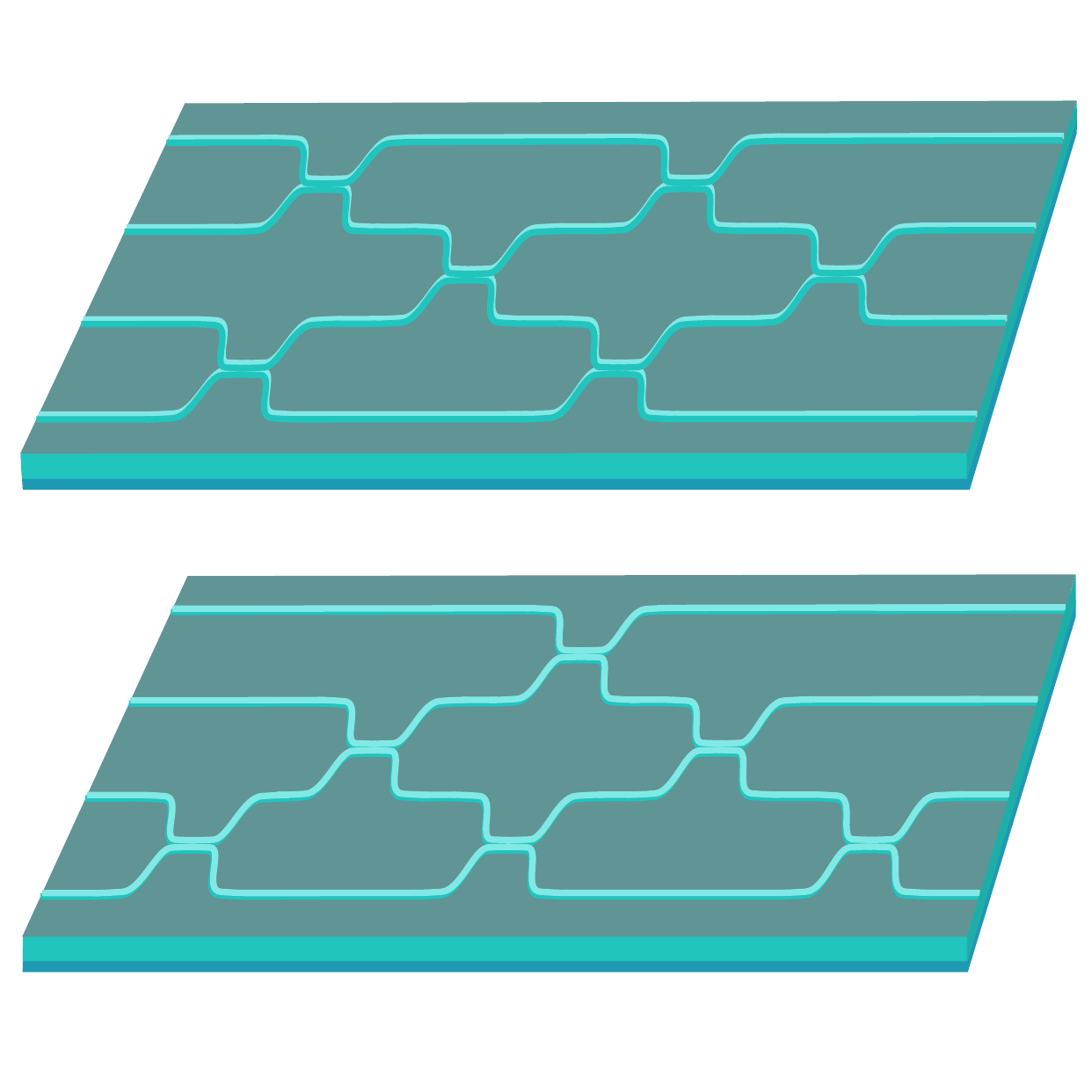} \label{subfig:mesh}}
    \\
    \subfloat[]
    {\includegraphics[width=0.85\textwidth]{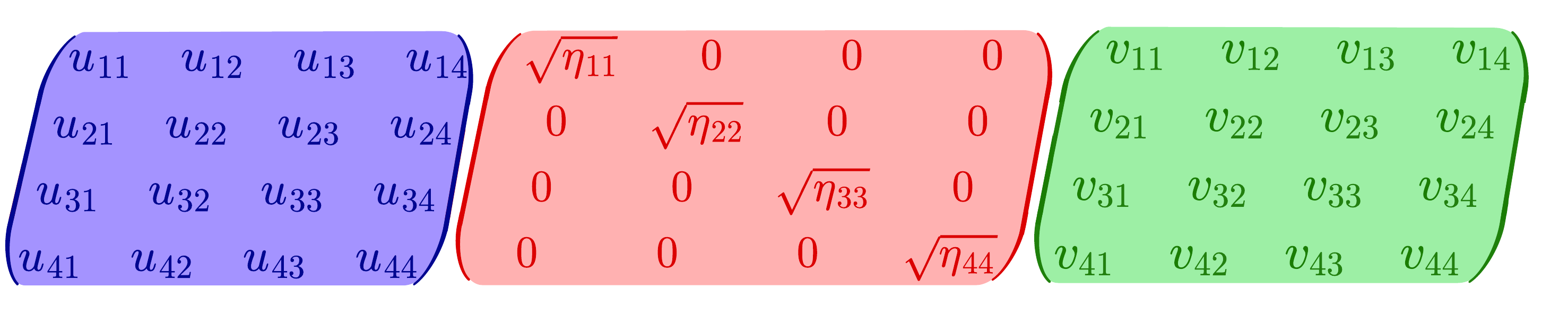} \label{subfig:SVD}}
    \caption{\label{fig:pipeline} The overall pipeline to describe the optical losses of a device is streamlined. (a) First, the unitary matrix is sampled according to the Haar measure, granting a uniform distribution on the $n$-sphere starting from a fixed basis state -- such as the north pole on the 2-sphere. The 2-sphere here is a figurative example, the dimension is arbitrary. (b) The matrix is encoded into a \gls{mzi} mesh (or an equivalent device). The upper mesh depicts the Clement scheme, while the mesh at the bottom the Reck one. Here the losses are inserted per each \gls{mzi}, see Equation \eqref{eq:lossy_interf}. (c) The non-unitary matrix (due to the losses) is eventually decomposed through the \gls{svd}, see Equation \eqref{eq:SVD_losses}.}
\end{figure*}

\subsection{The bosonic basis}

In the first place, we introduce the variables acting in the phase space of the system. A linear $M$-modes multiport interferometer is defined as a unitary operator $\hat U$ acting on the ladder operators as follows~\cite[pag.33]{kok2010introduction}:
\begin{equation}
\label{eq:boson_interf}
    \hat U \hat a_i \hat U^\dagger = \sum_{j=1}^M U_{ij} \hat a_j
\end{equation}
The $U$ matrix can be decomposed in multiple ways, the most standard being into $2 \times 2$ operators, the \glspl{mzi}, which can implement any $U(2)$ operator. In the formalism we adopt, the \gls{mzi} is described as~\cite{kondratyev2020multiport,Clements}
\begin{equation}
\label{eq:interferometer}
    T_{i, i+1}(\theta, \phi) = 
    \begin{pmatrix}
        e^{i\phi} \cos(\theta) & -\sin(\theta) \\
        e^{i \phi} \sin(\theta) & \cos(\theta)
    \end{pmatrix}
\end{equation}
i.e. the action of a phase-shifter acting on the $i$-th mode followed by a beam-splitter acting over the $i$, $i+1$ modes.
Gaussian states can be easily described in the phase space, rather than in the Fock space.
In this frame, the phase space is characterized by $2M$ variables, namely $\hat q_i$ and $\hat p_i$, i.e. the canonical observables, with $i = 1, ..., M$ and $M$ the number of modes. This basis is referred to as Heisenberg basis, and it derives from a linear combination of the ladder operators~\cite{olivares2012quantum,brunelli2015single,beggi2018back}:
\begin{equation}
    \hat q = \frac{\hat a + \hat a^\dagger}{\sqrt 2}, \quad 
    \hat p = \frac{\hat a - \hat a^\dagger}{i\sqrt 2}
\end{equation}
In a more general picture, the Heisenberg basis is given by $\{\hat q_i, \hat p_i \}_{i=1}^M$, or $\hat{\textbf{r}} = (\hat{\textbf{q}}, \hat{\textbf{p}})$.
The basis corresponding to the ladder operators in the phase space, instead, is labelled as the bosonic basis, and it is described by the ladder operators $\{\hat a_i, \hat a^\dagger \}_{i=1}^M$, or $(\hat{\bm{a}}, \hat{\bm{a}}^\dagger)$.
Referring to Equation \eqref{eq:boson_interf}, the action of the multiport interferometer can be split between the $\hat a$ and $\hat a^\dagger$ operators:
\begin{equation}
\label{eq:U_symplectic}
    U \to \mathcal U =
    \begin{pmatrix}
        U & \mathbb O \\
        \mathbb O & U^*
    \end{pmatrix}
\end{equation}
with $U$ acting over $\hat a$ and $U^*$ over $\hat a^\dagger$. It is straightforward to prove that $\mathcal U$ is a symplectic matrix\color{black}, just check $\mathcal U^T \Omega \mathcal U = \Omega$, with $\Omega = \begin{pmatrix}
    \mathbb O & I_M \\ -I_M & \mathbb O
\end{pmatrix}$. \color{black}
Within this formulation, the matrix $\mathcal U$ determines the global transformation of the mesh-based photonic circuit \cite{ferraro2005gaussian,hamilton2017gaussian}.

\subsection{Squeezing transformations}

Another crucial transformation in the frame of Gaussian variables is represented by the squeezing operator $\hat S(r)$. Squeezing compresses the uncertainty over a single quadrature at the cost of increasing the noise over the conjugated variable, keeping the Heisenberg principle satisfied~\cite{lawrie2019quantum,suerra2026generation}.
The algebraic action of the squeezing on the ladder operator is described as~\cite{christ2014theory}
\begin{subequations}
\label{eq:squeezing}
\begin{align}
    \hat a_i \xmapsto{\hat S(r)} \hat a_i \cosh(r) + \hat a_i^\dagger \sinh(r)\\
    \hat a_i^\dagger \xmapsto{\hat S(r)} \hat a_i \sinh(r) - \hat a_i^\dagger \cosh(r)
\end{align}
\end{subequations}
Defining $\Sigma$ as the squeezing matrix acting over the bosonic basis $(\bm a, \bm a^\dagger)$, as $U$ from Equation \eqref{eq:boson_interf} for the interferometric circuit, its action on a single mode, i.e. on $(\hat a, \hat a^\dagger)$ can be shaped as
\begin{equation}
    \Sigma_2 = \begin{pmatrix}
        \cosh(r) & \sinh(r) \\
        \sinh(r) & \cosh(r)
    \end{pmatrix}
\end{equation}
For the general $M$ mode case, it can be described as~\cite{hamilton2017gaussian}
\begin{equation}
\label{eq:sigma_squeezing}
    \Sigma = \begin{pmatrix}
        \bigoplus_{i=1}^M \cosh(r_i) & \bigoplus_{i=1}^M \sinh(r_i) \\
        \bigoplus_{i=1}^M \sinh(r_i) & \bigoplus_{i=1}^M \cosh(r_i) 
    \end{pmatrix}
\end{equation}
$r_i$ being the squeezing factor for each $i$-th mode.
It is straightforward to prove that $\Sigma$ belongs to the symplectic group as well.
In this paper, we do not address the noise and optical losses arising from squeezed states. However, a description of these states is mandatory to describe how the covariance matrix is affected by optical losses from multiport interferometers.

\subsection{Transformations on the first and second moments}

In this Section we introduce the description of the first and second moments related to the variables introduced in the phase space. When dealing with Gaussian variables, the first and second moments suffice to describe the dynamics of the system. 
In the bosonic basis, the mean vector corresponds to the expectation value over the ladder operators:
\begin{equation}
    \bm \mu = (\bm \alpha, \bm \alpha^*)
\end{equation}
with $\alpha_i = \bra{\psi}\hat a \ket{\psi}$.
We introduce the second moment, or covariance matrix of the quantum system, as~\cite{hamilton2017gaussian}
\begin{equation}
\label{eq:cov_mat}
    \sigma_{ij} = \frac{1}{2} \bra{\psi} \{\hat a_i; \hat a_j^\dagger \}\ket{\psi} - \bra{\psi} \hat a_i \rangle \langle \hat a_j^\dagger \ket{\psi}
\end{equation}
with $\{ \; \, ; \; \}$ being the anti-commutator. \color{black}In the bosonic basis, the covariance matrix defines the correlations between the ladder operators belonging to different modes, as reported in the seminal paper on \gls{gbs} by Hamilton et al.~\cite{hamilton2017gaussian}. \color{black}
We have seen from Equation \eqref{eq:boson_interf} how the interferometric multiport acts linearly on the ladder operators and, from Equations \eqref{eq:squeezing}, how squeezing acts linearly as well. When these transformations act on the ladder modes, calling the global operator $\hat S$, we have $\ket{\psi} \to \hat S \ket{\psi}$, which affects the covariance matrix as follows:
\begin{subequations}
\label{eq:sympl_transf}
\begin{align}
    \bra{\psi} \hat S^\dagger \hat a_i \hat S \ket{\psi} = S_{ij} \bra{\psi} \hat S^\dagger \hat a_j \hat S \ket{\psi} \\
    \bra{\psi} \hat S^\dagger \hat a_i \hat a^\dagger_i \hat S \ket{\psi} = \\
    \nonumber
    \bra{\psi} \hat S^\dagger \hat a_i \hat S \hat S^\dagger \hat a^\dagger_i \hat S \ket{\psi} = 
    S_{ik} \bra{\psi} \hat a_k \hat a^\dagger_l \ket{\psi} S_{lj}^*
\end{align}
\end{subequations}
For the sake of streamlining the notation, we have introduced the Einstein convention over the repeated indices. Replacing the transformations in Equations \eqref{eq:sympl_transf} in the expression from Equation \eqref{eq:cov_mat}, we can deduce the general transformation $\sigma$:
\begin{equation}
\label{eq:sigma_evo}
    \sigma \mapsto S \sigma S^\dagger
\end{equation}

\subsection{\label{sec:att_ch}The attenuation channel}

In this Section, we describe photon losses as the interaction between the physical modes and the environment. 
Let us now turn to CP-maps apt to describe the effect of quantum noise and decoherence, from the coupling of the physical system with the environment. It is known in literature~\cite[pag.97]{serafini2023quantum} that the open dynamics of a Gaussian system can be described by the CP-maps:
\begin{equation}
    \sigma \xmapsto{CP} X \sigma X^T + Y \; , \quad Y + i \Omega \geq i X \Omega X^T\\[0.1cm]
\end{equation}
where
\begin{equation}
    \quad \Omega = \bigoplus_{j=1}^M\begin{pmatrix}
        0 & 1 \\ -1 & 0
    \end{pmatrix}\,\,.
\end{equation}
For a single physical mode, optical losses can be modelled by the attenuator channel, a deterministic Gaussian CP-map with
\begin{equation}
X = \cos\theta\,{I}_2 \; , \quad Y = (\sin\theta)^2\,\frac{{I}_2}{2}
\label{XY_lossy}
\end{equation}
${I}_2$ being the identity $2\times 2$.
In fact, we can assume the lossy channel as a beam-splitter operation, splitting the signal to external auxiliary modes. For a single mode, this operation assumes the form
\begin{equation}
\label{eq:rot_loss}
R_\theta =
\begin{pmatrix}
    \cos\theta\,{I}_2 &\sin\theta\,{I}_2\\
    -\sin\theta\,{I}_2 &\cos\theta\,{I}_2
\end{pmatrix}\,\,,
\end{equation}
Here the $2$-nd mode is set as a vacuum state, thus with covariance matrix $ I_2/2$.
The action of the lossy channel on the system with covariance matrix $\sigma$, interacting with the external environment with covariance matrix ${I}_2/2$, is represented by
\begin{align}
\nonumber
    &R_\theta\Big[\frac{{I}_2}{2}\oplus\sigma\Big]R^{\mathsf{T}}_\theta = \\
    \nonumber
    &\begin{pmatrix}
        \cos\theta\,{I}_2 & \sin\theta\,{I}_2 \\
        - \sin\theta\,{I}_2 & \cos\theta\,{I}_2
    \end{pmatrix}
    \begin{pmatrix}
        \frac{{I}_2}{2}& \mathbb O \\
        \mathbb O & \sigma
    \end{pmatrix}
    \begin{pmatrix}
        \cos\theta\,{I}_2 & -\sin\theta\,{I}_2 \\
        \sin\theta\,{I}_2 & \cos\theta\,{I}_2
    \end{pmatrix}=\\
    &\begin{pmatrix}
    (\sin\theta)^2\sigma+(\cos\theta)^2\frac{{I}_2}{2} &\sin\theta\cos\theta(\frac{{I}_2}{2}-\sigma)\\
        \sin\theta\cos\theta(\frac{{I}_2}{2}-\sigma) &(\cos\theta)^2\sigma+(\sin\theta)^2\frac{{I}_2}{2}
    \end{pmatrix}
    \label{final_cov_matrix}
\end{align}
So the partial trace over the environment of the final covariance matrix (\ref{final_cov_matrix}) is obtained by pinching the lower-right $2\times2$ block, which corresponds to the system variables:
\begin{equation}
\sigma \longmapsto
(\cos\theta)^2\sigma+(\sin\theta)^2\frac{{I}_2}{2}\,\,.
\end{equation}
Overall, this dynamics just corresponds to the action of the deterministic Gaussian channel with matrices as in (\ref{XY_lossy}).
regarding the loss parameter as $\cos(\theta)=\sqrt \eta$, and therefore $\sin(\theta) = \sqrt{1-\eta}$, the covariance matrix of the single-mode physical system subject to optical losses may be written as
\begin{equation}
\label{eq:sigmaSimple}
    \eta\,\sigma + (1-\eta) \frac{{I}_2}{2}\,\,.
\end{equation}
This form is reported by many works~\cite{nehra2026all,oh2024classical} which introduce optical losses when describing the covariance matrix of the system.

\subsection{From experimental data to a theoretical frame}

As aforementioned in the previous Section, the lossy channel can be represented as a beam-splitter pumping out a certain amount of probability amplitudes from the physical modes to the environmental ones. The optical loss due to the passage of photons in an interferometer can be measured, and in general it consists of three steps, i.e. the passage of light through the coupler (edge coupler or rather grating couplers) into the integrated chip, the interaction of its beams through the interferometer and, eventually, the output, which involves another passage through the coupler to the external mean (in general fiber, or even air in a bulk implementation). Each passage involves a certain amount of losses, which can be measured directly e.g. with coherent light. Accounting for all the previous interfaces, a $2 \times 2$ mesh, consisting of a single interferometer, results in the following configuration:
\begin{align}
\label{eq:lossy_interf}
\nonumber
    &T_{\bm{\eta}}(\theta, \phi) = \\
    &\begin{pmatrix}
        \sqrt \eta_3 & 0 \\
        0 & \sqrt \eta_4
    \end{pmatrix}
    \begin{pmatrix}
        \sqrt \eta_{bs} & 0 \\
        0 & \sqrt \eta_{bs}
    \end{pmatrix}\,
    T(\theta,\phi)\,
    \begin{pmatrix}
        \sqrt \eta_1 & 0 \\
        0 & \sqrt \eta_2
    \end{pmatrix}
\end{align}
In this picture, $T_{\bm\eta}$ represents the lossy interferometer, while $T(\theta, \phi)$ the logical operation to be implemented.
The $\eta_1$, $\eta_2$ stand for the losses due to the interface when the photons enter the beam-splitter, $\eta_{bs}$ are the losses induced by the beam-splitter itself, $\eta_3$, $\eta_4$ on the contrary are the losses related to the interface when leaving the beam-splitter. 
Take notice, the formulation of $T_{\bm \eta}$ strictly depends on the specifics of the hardware, e.g. material, computational architecture or packaging.
\color{black}
From a theoretical perspective, each lossy channel can be considered as an interaction between the physical mode, suffering from optical loss, and an external ancillary mode, as described in the scheme of the attenuator channel in Section \ref{sec:att_ch}. Such an interaction can be characterized as a beam-splitter, following the formalism in Equation \eqref{eq:rot_loss}:
\begin{equation}
\begin{split}
    &\begin{pmatrix}
    \sqrt{\eta}_i & 0 \\
        0 & \sqrt{\eta_{i+1}}
    \end{pmatrix} \\ \to
    &\begin{pmatrix}
        \sqrt{\eta_i} & 0 & \sqrt{1 - \eta_i} & 0 \\
        0 & \sqrt{\eta_{i+1}} & 0 & \sqrt{1-\eta_{i+1}} \\
        -\sqrt{1 - \eta_i} & 0 & \sqrt{\eta_i} & 0 \\
        0 & -\sqrt{1-\eta_{i+1}} & 0 &\sqrt{\eta_{i+1}}
    \end{pmatrix}
\end{split}
\end{equation}
where $\sqrt{\eta_i} = \cos(\theta_i)$ and, consequently, $\sqrt{1-\eta_i} = \sin(\theta_i)$. If acting on both the ladder operators $\hat a_i$, $\hat a_i^\dagger$, we would insert the $I_2$ identity in front of each $\eta_i$, i.e. $\sqrt{\eta_i} \to \sqrt{\eta_i} I_2$ and $\sqrt{1-\eta_i} \to \sqrt{1-\eta_i} \, I_2$, achieving the same transformation (in the above example, for two modes instead of a single one) in Section \ref{sec:att_ch}.
\color{black}
However, within this scheme, each time a new $\eta_i$ is introduced, a new external channel is consequently introduced as well, which could polynomially raise the number of external modes if we implement a mesh of \glspl{mzi} with a huge number of imperfect interferometers $T_{\bm\eta}$.
To avoid such a trend, the \gls{svd} can be leveraged. In the first place, we construct the non-unitary matrix $A$ arising from the mesh of imperfect interferometers $T_{\bm \eta_i}$:
\begin{equation}
    A = \prod_{i=1}^{M(M-1)/2}T_{\bm\eta_i}
\end{equation}
$M(M-1)/2$ being the number of \glspl{mzi} in the mesh, while $A$ represents the no-more unitary operator acting over the modes of the system. $A$ describes the action of the imperfect multiport interferometer over the annihilation operators $\hat a_i$, while the action over the creators is still represented by $A^*$. The overall operator can be thus described as $\mathcal A = A \oplus A^*$.

\subsection{Exploiting the SVD}

A common technique used to study a lossy mesh of \glspl{mzi} leverages the \gls{svd}. In this frame, all the optical losses are gather in the mid of the circuit, and the $\mathcal A$ operator can be reduced to
\begin{equation}
\label{eq:SVD_losses}
    \mathcal{A} = \mathcal U_2 \sqrt{H} \mathcal U_1
\end{equation}
While $\mathcal U_1$, $\mathcal U_2$ are still unitary matrices, the diagonal matrix
\begin{equation}
\label{eq:sqrt_H}
\sqrt{H} = \mathrm{diag}(\sqrt{\eta_1}, ..., \sqrt{\eta_{2M}})
\end{equation}
now contains all the dissipative factors $\eta_i$, i.e. the total amount of attenuation per channel. From this point onward, we refer to $\sqrt{H}$ as the loss matrix: if $\eta_i=1 \; \forall i$ no loss occurs, while $\eta_i=0 \; \forall i$ means total dissipation. Take notice, since $A$ and $A^*$ share the same singular values (the singular value being real), in the loss matrix $\sqrt H$ the diagonal coefficients $\sqrt{\eta_{i}}$ are repeated pairwise.

\section{\label{sec:cov_mat}Covariance matrix of a lossy interferometry}

In this Section, we prove how to describe the covariance matrix of the system in a \gls{gbs} process by composing a mesh of imperfect interferometers such as from Equation \eqref{eq:lossy_interf}.
\newline
\newline
\textbf{Theorem}
The covariance matrix $\sigma_S$ of a \gls{gbs} performed with imperfect interferometers is given by
\begin{equation}
    \sigma_S = \frac{1}{2} (I_{2M} - \mathcal{U}_2 H \mathcal U_1^\dagger + \mathcal U_2 \sqrt H \mathcal U_1 \Sigma^2 \mathcal U_1 \sqrt H \mathcal U_2^\dagger)
\end{equation}
$I_{2M}$ being the identity matrix, $\Sigma$ the matrix implementing the squeezing on the ladder operators and $\mathcal U_1$, $\mathcal U_2$, $\sqrt H$ the components from the \gls{svd} in Equation \eqref{eq:SVD_losses}.
\newline
\newline
\textbf{Proof}
Since each lossy channel can be described as the CP map in Section \ref{sec:att_ch}, the attenuation factors introduced by the loss matrix $\sqrt H$ can be described as the action of a beam-splitter between the modes of the system $S$ and the external, environmental modes $E$:
\begin{equation}
    \sqrt H \to 
    \begin{pmatrix}
        \sqrt H & \sqrt{I_{2M} - H} \\
        -\sqrt{I_{2M} - H} & \sqrt H
    \end{pmatrix}
\end{equation}
The matrix introduced above accounts for a size $4M \times 4M$. The operators $\mathcal U_i$, due to their unitarity, do not act on the environmental modes. The overall $\mathcal A$ operator from Equation \eqref{eq:SVD_losses} can now be embedded into the $4M \times 4M$ unitary group:
\begin{equation}
\label{eq:U_losses}
    \mathcal{A} = 
    \begin{pmatrix}
     I_{2M} & \mathbb O \\
    \mathbb O & \mathcal U_2 
    \end{pmatrix}
    \begin{pmatrix}
    \sqrt{H} & \sqrt{ I_{2M} -H} \\
     -\sqrt{ I_{2M} -H} & \sqrt{H} 
    \end{pmatrix}
    \begin{pmatrix}
     I_{2M} &  \mathbb O \\
    \mathbb O & \mathcal U_1
    \end{pmatrix}
\end{equation}
The matrix $\mathcal{A}$ now belongs back to a higher dimensional unitary group, thanks to the interaction with the environment. Indeed, the upper-left block of the matrices in Equation (\ref{eq:U_losses}) involves the operation on the environment, while the lower-right block refers to the ideal photonic circuit. The off-diagonal blocks represent the interaction between the environment and the circuit.
The symplectic global operation now reads $S = \mathcal A \Sigma'$, with $\Sigma'$ being this time a $4M \times 4M$ matrix:
\begin{equation}
    \Sigma' = \begin{pmatrix}
        I_{2M} & \mathbb O \\
        \mathbb O & \Sigma
    \end{pmatrix}
\end{equation}
The lower block $\Sigma$ matrix refers to the original squeezing matrix from Equation \eqref{eq:sigma_squeezing}.
The global $4M \times 4M$ symplectic matrix assumes the form:
\begin{widetext}
\begin{equation}
\label{eq:symplectic_env}
    S =
    \begin{pmatrix}
        {I}_{2M} & \mathbb O \\
        \mathbb O & \mathcal U_2
    \end{pmatrix}
    \begin{pmatrix}
        \sqrt H & \sqrt{{I}_{2M}-H} \\
        -\sqrt{{I}_{2M}-H} & \sqrt H
    \end{pmatrix}
    \begin{pmatrix}
        {I}_{2M} & \mathbb{O} \\
        \mathbb{O} & \mathcal U_1
    \end{pmatrix}
    \begin{pmatrix}
        {I}_{2M} & \mathbb O \\
        \mathbb O & \Sigma
    \end{pmatrix}
\end{equation}
\end{widetext}
Referring to the evolution of the covariance matrix in Equation \eqref{eq:sigma_evo}, in a \gls{gbs} process the $M$ input modes are set in the vacuum state, so the covariance matrix for such state can be represented as $\sigma = {I}_{2M}/2$~\cite{simon1994quantum}, while its first moment as a null vector.
Thus, the evolution of the second moment of the states through an imperfect photonic circuit turns into
\begin{equation}
\label{eq:sigma_ideal}
    \sigma = \frac{1}{2} S S^{\dagger} = \frac{1}{2}
    \begin{pmatrix}
        \sigma_E & \sigma_{ES} \\
        \sigma_{ES}^\mathsf{T} & \sigma_S
    \end{pmatrix}
\end{equation}                                                                %
with $S$ always referring to Equation \eqref{eq:symplectic_env}
Since we are considering a joint system in a Gaussian state~\cite{brask2022gaussianstatesoperations}, with $E$ standing for the external environment, and $S$ for the physical system, the external part has to be traced out. The covariance matrix of the  physical system, $\sigma_S$, reads as
\begin{equation}
\label{eq:sigma_losses}
    \sigma_S = \frac{1}{2}\left(
        {I}_{2M} - \mathcal U_2 H \mathcal U_2^\dagger + \mathcal U_2 \sqrt H \mathcal U_1 \Sigma^2 \mathcal U_1^\dagger \sqrt H \mathcal U_2^\dagger \right)
\end{equation}
which ends the proof. $\blacksquare$
\newline
Thanks to such a formulation, the covariance matrix can be analytically estimated under a noisy regime related to the single device and its imperfections. To achieve this goal, the interferometers have to be properly characterized, see Equation \eqref{eq:lossy_interf}.
\newline
\newline
\textbf{Corollary 1}
If the loss matrix $\sqrt{H}$ is proportional to the identity, the formulation of $\sigma_S$ can be reduced to a multi-channel generalization of uniform losses in Equation \eqref{eq:sigmaSimple}.
\newline
\textbf{Proof}
If the losses are adapted to a uniform channel, i.e. $H = \eta I$, such a statement makes $\sigma_S$ as
\begin{align}
\nonumber
\sigma_S &= \frac{1}{2}\left({I}_{2M} - \eta\,\mathcal U_2 \mathcal U_2^\dagger + \eta\,\mathcal U_2 \mathcal U_1 \Sigma^2 \mathcal U_1^\dagger \mathcal U_2^\dagger \right)
\end{align}
Here $\mathcal U_2 \mathcal U_1 = \mathcal U$, i.e. the unitary operator implementing the mesh of \glspl{mzi}, and $\mathcal U \Sigma = S$, i.e. the global symplectic transformation. The covariance matrix now reads
\begin{align}
\sigma_S =\frac{{I}_{2M}}{2} (1 - \eta) + \eta \sigma
\end{align}
with $\sigma = S S^\dagger/2$, matching the expression reported in Equation \eqref{eq:sigmaSimple}. $\blacksquare$
\newline
\newline
\textbf{Corollary 2}
If no squeezing occurs, $\sigma_S$ corresponds to the covariance matrix of the vacuum state.
\newline
\textbf{Proof}
If no squeezed light is injected into the \gls{mzi} mesh, $\Sigma = I_{2M}$, and therefore the expression in Equation \eqref{eq:sigma_losses} becomes
\begin{equation}
    \sigma_S = \frac{1}{2} (I_{2M} - \mathcal U_2 H \mathcal U_2^\dagger + \mathcal U_2 \sqrt{H} \mathcal U_1 \mathcal U_1^\dagger \sqrt H \mathcal U_2^\dagger ) = \frac{I_{2M}}{2}
\end{equation}
which corresponds to the covariance matrix of the vacuum state. $\blacksquare$
\newline
\newline
\textbf{Corollary 3}
In case of total dissipation, the sole non-null outcome from a \gls{gbs} process is counting no photons.
\newline
\textbf{Proof}
If we consider a complete dissipation process, i.e. no output photons, we have to set $H = \mathbb O$. \gls{gbs} is regarded as the paramount algorithm which leads to quantum supremacy in the Gaussian framework. The core of \gls{gbs} lies in exploiting the covariance matrix and sampling the photon counts to compute the Hafnian of a matrix $\mathcal A$, a \#P task whose complexity scales exponentially with the size of the system~\cite{bjorklund2019faster}. The formulation and the notation of this algorithm is summarized in Appendix \ref{app:GBS}.

Referring to Equation \eqref{eq:sigma_losses}, the covariance matrix of the system thus reads
\begin{equation}
    \sigma = \frac{{I}_{2M}}{2}
\end{equation}
which matches with the covariance of the vacuum state.
When performing \gls{gbs}, the $Q$ matrix becomes
\begin{equation}
    Q = \sigma + \frac{{I}_{2M}}{2} = I_{2M}
\end{equation}
thus the $A$ matrix reads
\begin{equation}
\mathcal{A} = X_{2M} ({I}_{2M} - Q^{-1}) = \mathbb O
\end{equation}
When we compute the probability to count zero output photons, the analytical expression returns
\begin{equation}
\label{eq:prob_haf}
P(\bm{0}) = \frac{1}{\bm{0}! \sqrt{\text{Det}(Q)}} \mathrm{Haf}([\,]) = 1
\end{equation}
$[\,]$ being a matrix with size $0 \times 0$, whose Hafnian conventionally returns $1$, as well as $0!=1$.
Any case accounting for a number of samples different from zero happens with null probability, since the Hafnian of the null matrix $\mathbb O$ returns a value equal to zero. $\blacksquare$

\section{\label{sec:numerical}Numerical simulations}

In this Section, we showcase some numerical examples of photonic simulations run on a Gaussian backend, relying on the formalism introduced in the previous Section, from which we can simulate an ideal or a lossy \gls{gbs} process.

The main pipeline of these experiments consists first of all of setting a squeezing parameter $r$, and sampling a random unitary matrix $U$ with size $M \times M$ exploiting the Haar measurement, see Appendix \ref{app:Haar}. From here we build the ideal interferometric symplectic matrix (\ref{eq:U_symplectic}), and the squeezing symplectic matrix (\ref{eq:asqueezing}), and then evaluate the covariance matrix $\sigma$ of the output state, and extract the output distribution computing Equation (\ref{pr_gbs}). In order to simulate the lossy \gls{gbs} process instead, we decompose the random unitary matrix $U$ into a composition of ideal \gls{mzi} matrices via Reck or Clements schemes, as explained in Appendix \ref{app:Reck}, and then add a fixed loss factor $\eta$ to each interferometer, so we can perform the \gls{svd} decomposition presented in Equation (\ref{eq:SVD_losses}) to evaluate the loss matrix $\sqrt{H}$, and thus also the covariance matrix $\sigma_S$ of the photonic circuit (\ref{eq:sigma_losses}) subjected to photon loss. Then in the same way as before, we can extract the output distribution. Via this pipeline, we are able to simulate a noisy (non-Gaussian) \gls{gbs} process, and examine the loss matrix which give rise to a certain output distribution. This overall scheme is reported in Figure \ref{fig:pipeline}.

In applying the loss parameter onto a \gls{mzi} matrix, we discard the optical losses due to the packaging, i.e. when light enters and exits the chip, and we focus on the sole terms depending on the imperfections of the \glspl{mzi}. In fact, while the optical losses due to the interferometric components are ubiquitous in any photonic device, packaging may drastically differ depending on the considered architecture (e.g. loop-based versus spatial-bin encoding). Referring to Equation \eqref{eq:lossy_interf}, the lossy interferometer acting on the $j$, $j+1$ modes, with no losses due to the packaging, can be described as
\small
\begin{align}
\label{eq:loss_interf_big}
\nonumber
    &T_{\bm\eta}(\theta, \phi) = \\
    &=\begin{pmatrix}
        \ddots & & & & \\
         & 1 & 0 & 0 & 0 \\
         & 0 & \sqrt{\eta_{bs}} & 0 & 0 \\
         & 0 & 0 & \sqrt{\eta_{bs}} & 0 \\
         & 0 & 0 & 0 & 1 \\
          & & & & & \ddots \\
    \end{pmatrix}
    \begin{pmatrix}
        \ddots & & & & \\
         & 1 & 0 & 0 & 0 \\
         & 0 & e^{i\phi}C_\theta & -S_\theta & 0 \\
         & 0 & e^{i\phi}S_\theta & C_\theta & 0 \\
         & 0 & 0 & 0 & 1 \\
          & & & & & \ddots \\
    \end{pmatrix}
\end{align}
%
where $C_\theta= \cos(\theta)$, $S_\theta = \sin(\theta)$. The non-diagonal elements are located in the $(j, j+1)$ and $(j+1,j)$ cells.
We set $\eta_{bs}=0.9$ for all the \glspl{mzi}, which allows to characterize optical losses for quite large matrices (up to $M=20$), otherwise we notice that the dissipation is too strong to reach high dimensions (the symplectic matrix resulting in being mostly null). We set uniform losses across the whole mesh, further characterizations can be performed regarding the specifications of the device and its architecture. In fact, two main approaches dominate the \gls{cv} photonic computation, spatial and time-bin encoding. The first approach relies on the Clement's scheme, since it makes the circuit more compact therefore attenuating optical dissipation. The time-bind encoding, on the contrary, is equivalent to the Reck scheme. These two decompositions are shown in Figure \ref{subfig:mesh}.

We perform two classes of numerical experiments. In the first one, we exploit our Gaussian backend to show how the probability distribution from a \gls{gbs} process is affected by losses. For the sake of simplicity, the numerical simulations are run on the sole Reck's setting.
In the second numerical experiment, we sample $n$ unitary matrices, distributed according to the Haar measure, we decompose them both into Clements and Reck schemes, and we assess in both these schemes how the loss matrix $H$ differs from the identity, i.e. how the losses are far from being uniform in the multiport interferometer.
As for the software, we rely on \href{https://pypi.org/project/pnn/}{PNN}, a Python library to decompose unitary matrices into a \gls{mzi} mesh according to both Clements and Reck schemes~\cite{liu2022reduce,liu2022reliability}. However, to decompose the generic unitary matrix $\mathcal U$ in the Reck's scheme, we relied on the algorithm explained in \ref{app:Reck}.
Moreover, in the Introduction as well as in the previous paragraph, we have mentioned the fact that the loop architecture is equivalent to the Reck scheme. This statement holds for the ideal case.
However, in the spatial and loop-based Reck configurations, the characterization of the losses slightly differs~\cite{qi2018linear}, an issue which is addressed in~\ref{app:Reck_loss}. Since the experiments carried out on the Reck scheme aim to simulate the dynamics in a loop architecture, we set the optical losses in the time-bin frame.

\subsection{GBS experiments in presence of losses}

In this Section, we compare the probability distribution outcomes of a \gls{gbs} in the ideal case and in a realistic one. As aforementioned, we set the loss factors $\eta=0.9$ for all the interferometers, adopting the Reck scheme with losses distributed according to the loop architecture. In all the numerical simulations, the squeezing factor has been tuned to $r=0.5$ for all modes. 
We highlight the fact that, exploiting a Gaussian backend in a simulative environment, no need of sampling is required -- differently from a real experiment -- since the probability distribution is directly provided by the Hafnian of $A$, see Equation \eqref{eq:A_def}. On a classical simulator, retrieving the probability related to high counts of photons is prohibitive, due to the computational time, thus we truncate the outcomes up to two photons without loosing of generality.
We settle two cases, picking up random angles in the first scenario for meshes of size $M(M-1)/2$ with $M=2,3,4,5$. Instead, in the second scenario, the angles are kept null, in order to characterize the full transmissivity of the device in presence of optical losses. Thus, in this second frame we are studying how $M$ non-entangled squeezed pure states are affected by such a noise.
\begin{figure}[h]
    \centering
    \includegraphics[width=1\linewidth]{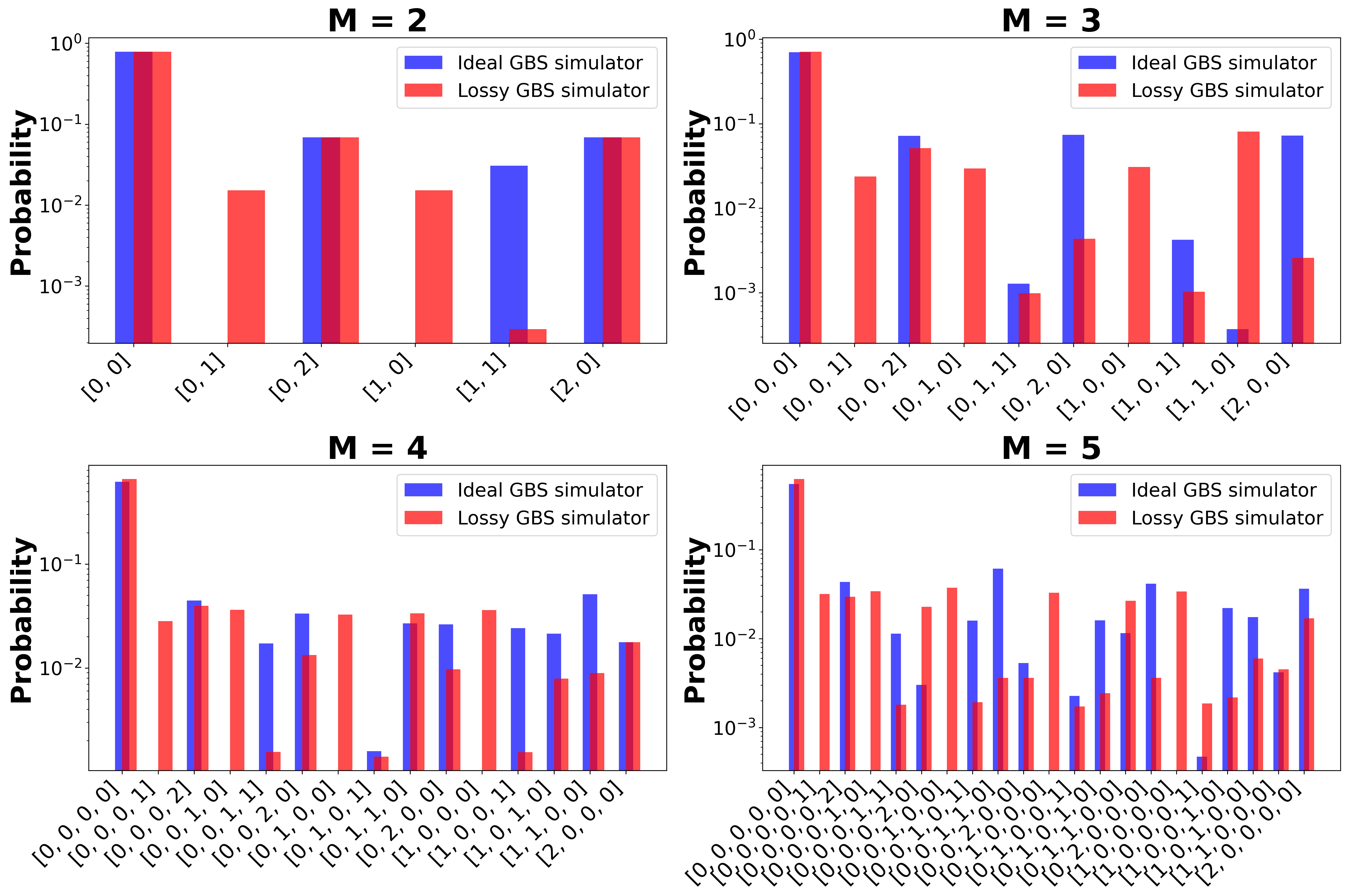}
    \caption{\label{fig:random_gbs}The comparison of ideal \gls{gbs} output distributions (in blue) along with the noisy ones (in red). The experiments have been repeated with different numbers of modes $M=2,3,4,5$. All the plots are reported in logarithmic scale. The \gls{mzi} parameters $\theta_i, \phi_i$ have been chosen randomly and fixed for every setting varying with $M$.}
\end{figure}
In Figure \ref{fig:random_gbs}, we compare the ideal distribution of a \gls{gbs} process along with a noisy one. The bins refer to the photon counts per channel. Just for instance, $[0,2,0]$ means that two photons have been detected on the second waveguide, while the first and third ones do not yield any count.
Since squeezed states are a superposition of occupation states with even numbers of photons, in the ideal case odd photon counts are not allowed. However, due to optical losses, these outcomes become accessible (e.g. $[0,1]$, $[0,1,0]$ and so on). The histogram is reported in a logarithmic scale, however, we can notice the following trend, with the vacuum state being more populated in the noisy scenario compared to the ideal one. States being more populated in the noisy scenario with respect to the ideal one inherit counts from higher-population states subject to losses.
\begin{figure}[h]
    \centering
    \includegraphics[width=1\linewidth]{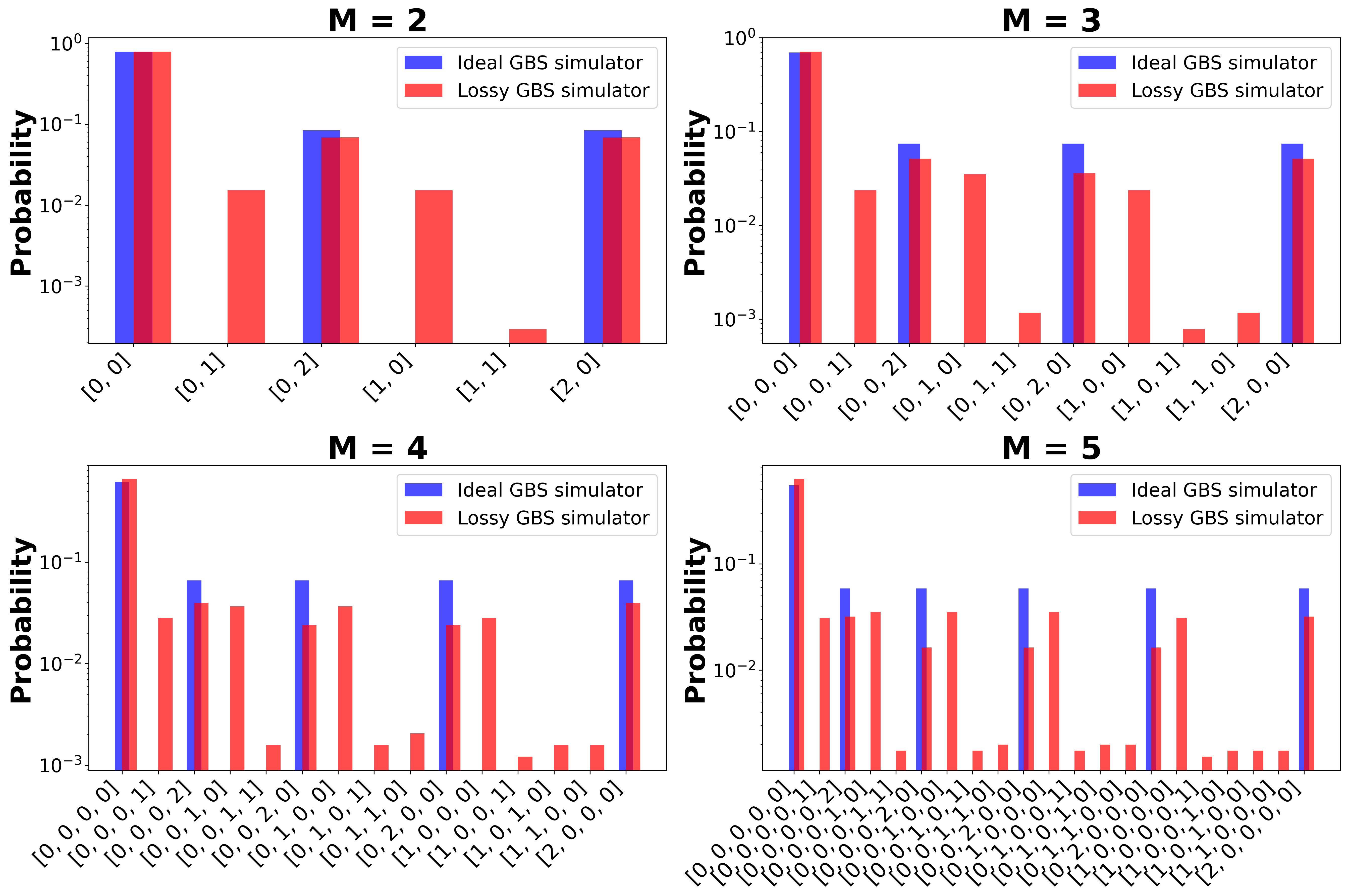}
    \caption{\label{fig:semi_gbs}The comparison of ideal \gls{gbs} output distributions (in blue) along with the noisy ones (in red). In this frame, the \glspl{mzi} are set as the identity, i.e. $\theta_i=\phi_i=0 \, \forall i$, in order to characterize how the counts from separable squeezed states are affected by losses. The experiments have been repeated with different numbers of modes $M=2,3,4,5$. All the plots are reported in logarithmic scale.}
\end{figure}
Instead, in Figure \ref{fig:semi_gbs}, we tune the angles of the mesh to be null. In fact, we want to compare the outcome distribution of counts from separable squeezed states in both the ideal and noisy cases. In the ideal case, equally populated states yield the same probability of being counted. Optical noise breaks such a symmetry. Again, the vacuum state in the lossy case tends to be more populated, as expected, rather than in the ideal frame.
All the parameters of the simulations are reported in Table \ref{tab:gbs_params}.
\begin{table}[t]
\centering
\caption{\label{tab:gbs_params}The parameters for the \gls{gbs} simulations, for both the experiments with random fixed and null mesh angles. Here $r$ represents the squeezing factor, $\eta$ the loss parameter adopted for all the \glspl{mzi}, $\gamma$ the maximum number of detected photons before the truncation in the simulation, and $M$ the number of modes for which the experiment has been repeated.}
\begin{tabular}{ll}
\toprule
\textbf{Hyperparameter} & \textbf{Value} \\
\hline
$r$ & $0.5$ \\ 
 $\eta$ & $0.9$ \\ 
 $\gamma$ & $2$ \\ 
 $M$ & $2,3,4,5$ \\
 \hline
\end{tabular}
\end{table}

\subsection{\label{sec:loss_matrix_num}Characterization of the loss matrix}

In this Section, we want to measure the distance, for fixed modes $M$ and multiple matrices sampled from the Haar measure, between the loss matrix $H$ introduced in Equation \eqref{eq:SVD_losses} and the identity $2M \times 2M$, i.e. to assess whether $H \propto I_{2M}$.
In many works~\cite{nehra2026all,oh2024classical}, this statement in fact is granted as an assumption, which allows to shift all the loss factors $\eta_i$ at the very beginning of the computation, a condition which simplifies the classical simulation of a \gls{gbs} process. As aforementioned in the Introduction, Ref.~\cite{su2020error} distinguishes two scenarios, i.e. Reck and Clements architectures, assuming that optical losses can be considered uniform only in the second scheme.
In Section \ref{sec:loss_matrix_num}, thanks to the formalism we developed up to now, we numerically check this statement.
To measure the effective distance between these two matrices, we first normalize both the identity and the loss matrix by their trace to fix this proportionality:
\begin{equation}
    \sqrt{H^{(l)}} \to \sqrt{\tilde H^{(l)}} = \frac{\sqrt{H^{(l)}}}{\mathrm{Tr}(\sqrt{H_l})} \; , \quad 
     I_{2M} \to \tilde{I}_{2M} = \frac{I_{2M}}{2M}
\end{equation}
Here $H^{(l)}$ represents the loss matrix associated with a specific symplectic matrix $S_l$, as in Equation \eqref{eq:decomp}, and, more importantly, with a sampled unitary matrix $U_l$ endowed with its losses.
Then we can compute the distance between these matrices as the trace norm~\cite{benenti2010computing}:
\begin{equation}
\label{eq:dist_matrix}
    d(\sqrt{\tilde H^{(l)}}, \tilde I) = \sqrt{\sum_{i=1}^{2M} \left(\sqrt{\tilde \eta_{ii}^{(l)}} - \frac{1}{2M}\right)^2}
\end{equation}
with $\sqrt{\tilde \eta_{ii}^{(l)}}$ being the diagonal elements belonging to $\sqrt{\tilde H_l}$. Moreover, if $\sqrt{\tilde \eta_{ii}} = \sqrt{\tilde \eta_{jj}} \; \forall i,j$, then trivially $\sqrt{\tilde \eta_{ii}} = 1/2M$ due to its normalization, and thus $d(\sqrt{\tilde H^{(l)}}, \tilde{I}_{2M}) = 0$.

\begin{figure}
    \centering
    \subfloat[]{\includegraphics[width=0.5\textwidth]{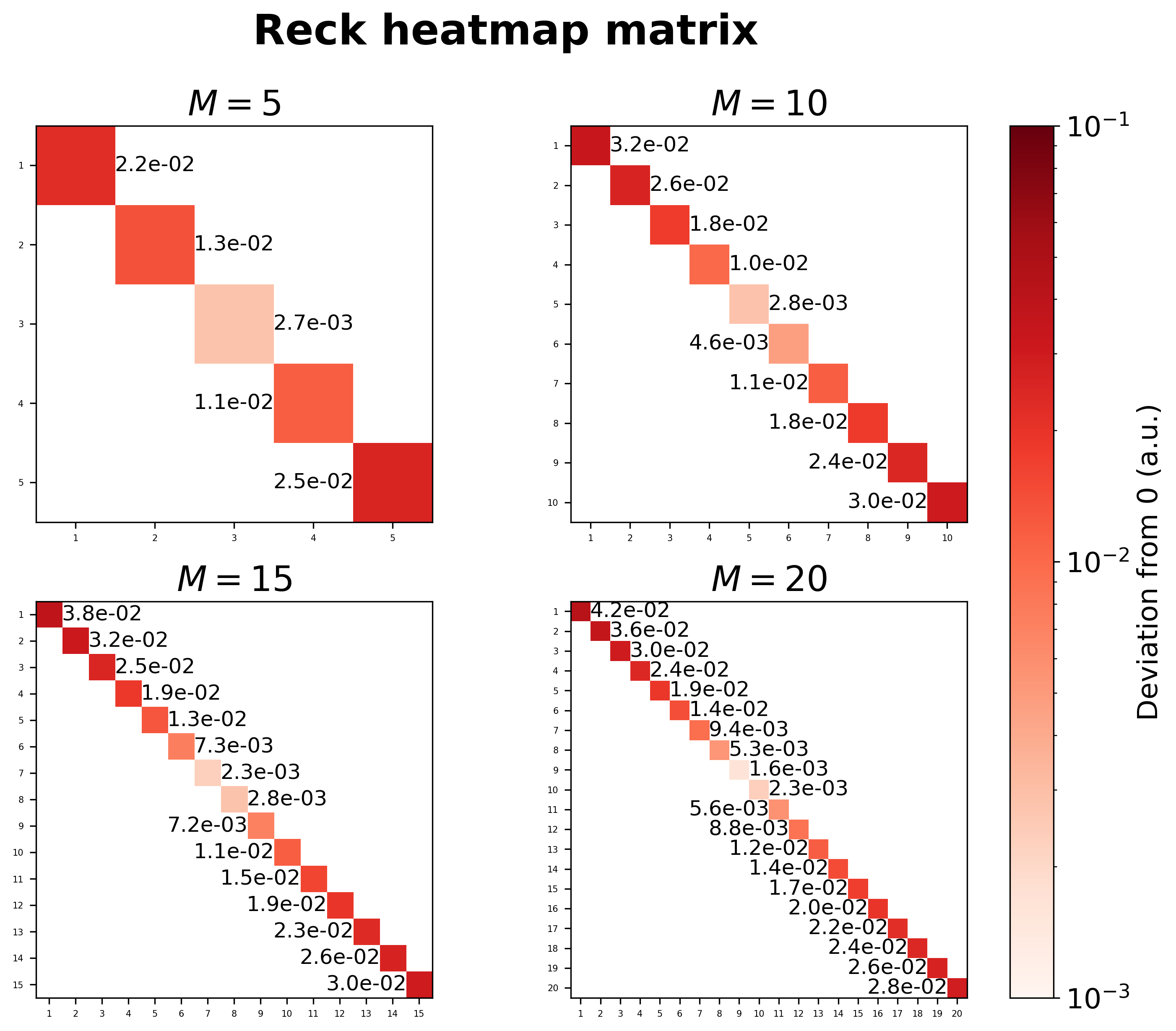}\label{subfig:heatmap_reck}}
    \\
    \subfloat[]
    {\includegraphics[width=0.5\textwidth]{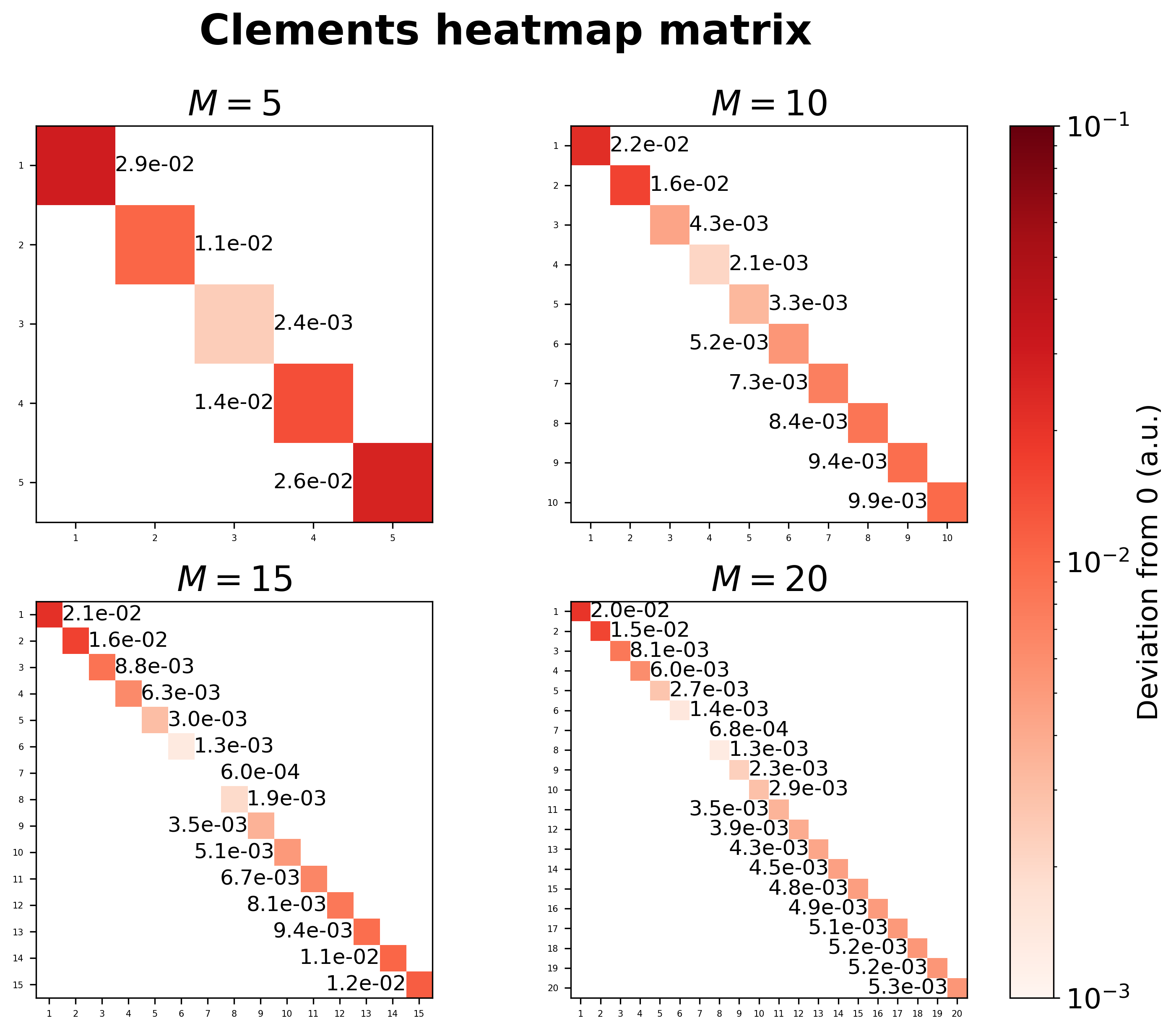} \label{subfig:heatmap_clements}}
    \caption{\label{fig:heatmaps} The heatmaps of the mean elementwise distances between the loss matrix and the identity, both normalized. Since the diagonal elements are pairwise equal, we half their number for the sake of readability. The distances are reported in \gls{au}.}
\end{figure}

Once the metric to compute the distance between the identity and the loss matrix has been assessed, we can focus on the numerical experiments. Following the scheme in Figure \ref{fig:pipeline}, we first sample $N$ random matrices according to the Haar distribution. These simulations have been repeated per each mode $M$, $M$ ranging from $2$ to $20$. The method for sampling the unitary matrices is detailed in \ref{app:Haar}. From the $M \times M$ unitary operator $U$ we then compute the operator $\mathcal U = U \oplus U^*$.
Once the set of $N$ unitary matrices is available, the next step consists of decomposing them into a specific mesh of interferometers. The experiments have been performed in both Reck and Clement schemes. The additional lossy terms in Equation \eqref{eq:loss_interf_big} are then inserted per each interferometer, yielding a global operator $2M \times 2M$ that, however, is no longer unitary, but in the form from Equation \eqref{eq:SVD_losses}. The loss matrix $\sqrt{H}$, and consequently $\sqrt{\tilde H}$, is retrieved performing the \gls{svd} on this non-unitary operator, calling the NumPy \texttt{linalg.svd} module. 
We show two numerical results. The first aims to show the heatmap of $\sqrt{\tilde H} - \tilde I$, i.e. of the distance between the matrices $\sqrt{\tilde H^{(l)}}$ and the normalized identity. To achieve this result, we first retrieve the $\sqrt{\tilde H^{(l)}}$ from their corresponding $N$ sampled unitary matrices, then we compute the following average per each component:
\begin{equation}
\label{eq:distances_coeff}
    \mu_i =  \sum_{l=1}^N \frac{\left|\sqrt{\tilde \eta^{(l)}_{ii}} - 1/2M\right|}{N}
\end{equation}
The vector $\boldsymbol \mu$ is reported in the heatmaps in Figure \ref{fig:heatmaps}, and it consists of the mean distances between all the diagonal coefficients of the loss matrices, labelled in Equation \eqref{eq:distances_coeff} as $\sqrt{\tilde \eta_{ii}^{(l)}}$ ($l$ referring to the $l$-th sample) and the normalized identity, i.e. $1/2M$. We showcase the heatmap for $M=5,10,15,20$. Since the matrix $\mathcal U = U \oplus U^*$ is the direct sum of two identical but conjugated matrices, the singular values from the \gls{svd}, conventionally ordered from the highest to the lowest values, are repeated pairwise equal. For the sake of illustration, we half their number in the heatmaps. The range of these values start from $10^{-1}$ to $10^{-4}$, being reported in Figure \ref{fig:heatmaps} in \gls{au}.
First of all, we compare these coefficient distances between the Reck scheme, Figure \ref{subfig:heatmap_reck}, and those from the Clements scheme, Figure \ref{subfig:heatmap_clements}. The latter ones show a lower discrepancy. In both schemes, the distances exhibit a minimum in the mid of the diagonal, meaning that the closer to the borders, the further from $1/2M$. However, in the Reck frame, the upper values increase with the number of modes $M$, while in the Clements picture the trend is the opposite. Instead, the lowest values oscillate in both cases.

\begin{figure}[h]
    \centering
    \includegraphics[width=1\linewidth]{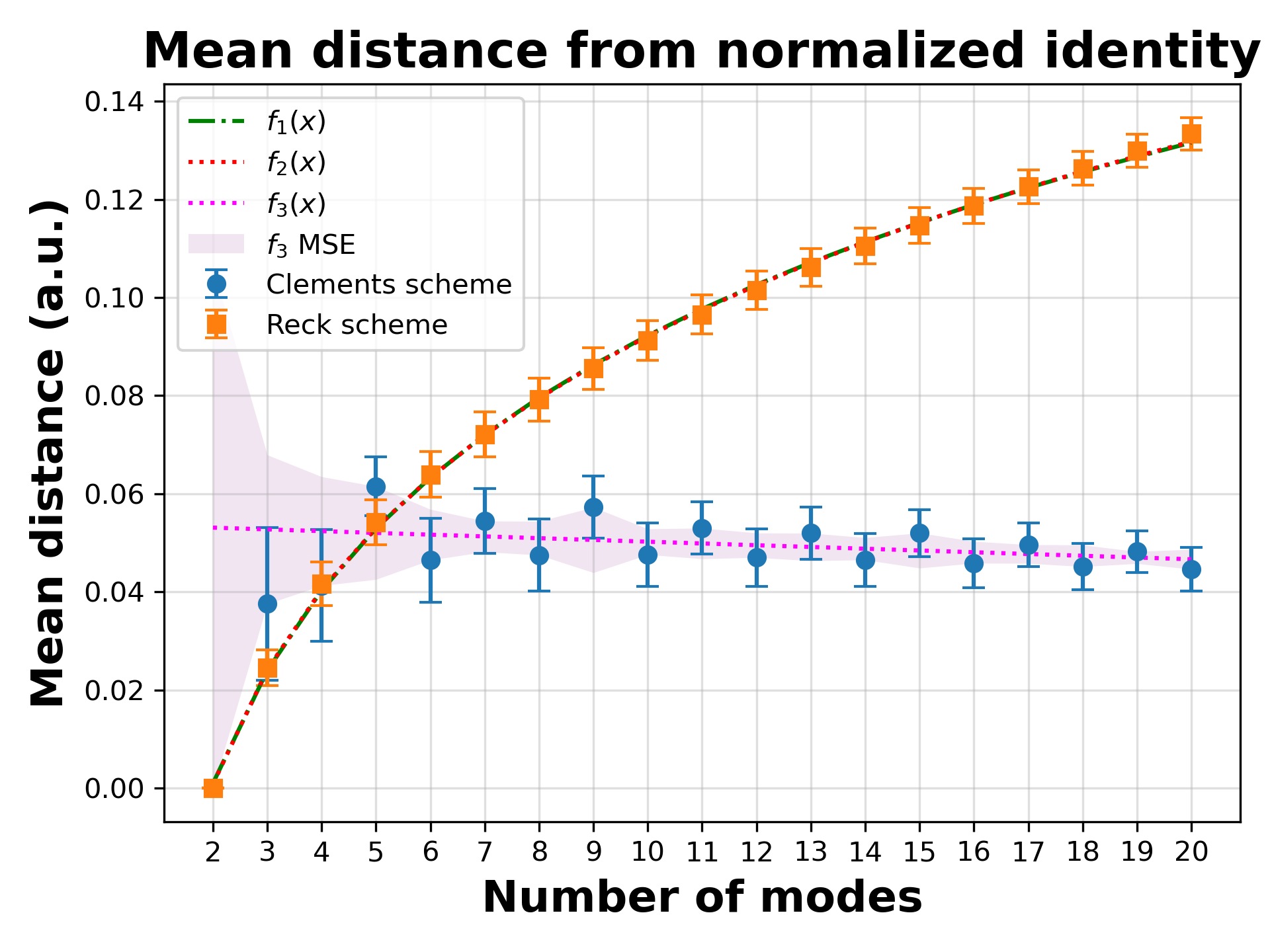}
    \caption{\label{fig:distances}The norm distances between the normalized loss matrix and the normalized identity, for both the Clements and Reck schemes. The trend is reported with respect to the number of modes. The functions $f_1$ and $f_2$ are defined in Equation \eqref{eq:fitting}\color{black}, while $f_3$ in Equation \eqref{eq:fittingf3}. The purple area represents the \gls{mse} evaluated at each single point for the $f_3(x)$ function.\color{black}}
\end{figure}

Eventually, as a last noteworthy aspect, we plot in Figure \ref{fig:distances} how the loss matrix diverge from the identity along with $M$. A straightforward consideration concerns the different trend between the two schemes. According to the elementwise distances in the heatmaps, in the Reck scheme the norm distance increases with $M$, while the in Clements frame the same distance tends to oscillate and converge into a bounded range \color{black}just after $M=5$\color{black}. Such a behaviour may derive from the fact that, in the loop-based Reck configuration, the losses are induced by ghost \glspl{mzi}, see Ref. \cite{qi2018linear} and Appendix \ref{app:Reck_loss}. 
As a last task, we try to assess the scaling law of the Reck divergence. We consider two candidates, a logarithmic trend $f_1$ and a power law $f_2$:
\begin{subequations}
\label{eq:fitting}
\begin{align}
    f_1(x) = a_1 + b_1 \log(x) \\
    f_2(x) = (x - a_2)^{b_2} + c_2
\end{align}
\end{subequations}
We deploy the \texttt{scipy.optimize} module to find the best parameters $a_1$, $b_1$ and $a_2$, $b_2$, $c_2$ to fit the curve of the Reck distance. Thereafter, we evaluated the \gls{mse} to compare the two results. We set $M=30$ to enhance the prediction of this analysis. The \gls{mse} related to the logarithm function $f_1$ is equal to $1.58 \times 10^{-4}$, while for the power law $f_2$ is around $4.51 \times 10^{-5}$.
\color{black}
Both $f_1$ and $f_2$ are fitting the curve compatibly with the error bars. Instead, we fit the Clements curve with a straight line:
\begin{equation}
\label{eq:fittingf3}
    f_3(x) = a_3 x + b_3
\end{equation}
Applying the same method as for the Reck case, we measured the \gls{mse} between the data points and $f_3$. We can see from Figure \ref{fig:distances} that the higher the number of modes, the lower the \gls{mse}, represented by the purple area. Again, $f_3(x)$ fits the data points up to one standard deviation.
\color{black}
%
%
All the parameters of the simulations are reported in Table \ref{tab:distance_params}.

\begin{table}[t]
\centering
\caption{\label{tab:distance_params}The parameters for the numerical simulations to assess the distance between the identity and the loss matrix. Here $\eta$ represents the loss parameter adopted for all the \glspl{mzi}, $M$ the range of the inquired modes, $N$ the number of unitary matrices sampled according to the Haar distribution per each mode $M$. The parameters $a_1$ and $b_1$ are optimized to make $f_1$ in Equation \eqref{eq:fitting} fit the curve of the Clements distance in Figure \ref{fig:distances}, while $a_2$, $b_2$, $c_2$ to make $f_2$ fit the same curve. The \gls{mse} is reported in the rows below.}
\begin{tabular}{ll}
\toprule
\textbf{Hyperparameter} & Va\textbf{}lue \\
\hline
$\eta$ & $0.9$ \\ 
 $M$ & $2 - 20$ \\ 
 $N$ & $10000$ \\
 $a_1$ & -0.043 \\
 $b_1$ & 0.059 \\
 $a_2$ & -0.357 \\
 $b_2$ & 0.055 \\
 $c_2$ & -1.045 \\
 $a_3$ & -$4 \times 10^{-4}$ \\
 $b_3$ & 0.0538 \\
 \gls{mse} $f_1$ & $1.58 \times 10^{-4}$ \\
 \gls{mse} $f_2$ & $4.51 \times 10^{-5}$ \\
 \gls{mse} $f_3$ & $3.4 \times 10^{-3}$ \\
 \hline
\end{tabular}
\end{table}

\color{black}
\subsection{Noisy sampling of the Hafnian}

The last numerical experiment consists of simulating a \gls{gbs} process to sample the Hafnian of a matrix. We benchmark the simulation of a noisy \gls{gbs} versus the hypothesis of optical losses set as uniform throughout all channels, in order to assess at what mode number $M$ and transmissivity $\eta$ the uniform-loss approximation diverges from the realistic one. To interpolate a valid loss matrix $\sqrt{H_{unif}}$ for a specific configuration of parameters $\{\theta_i; \phi_i \}$, it suffices to extract the value $\eta_{unif}$ which minimizes the distance between $\sqrt{H_{unif}} = \sqrt{\eta_{unif}} I_{2M}$ and the true loss matrix $\sqrt H$:
\begin{equation}
    \eta_{unif} = \mathrm{argmin}_{\eta} d\left(\sqrt \eta I_{2M}; \sqrt{H}\right)
\end{equation}
$d$ being the distance between diagonal matrices adopted from Equation \eqref{eq:dist_matrix}.
From the above equation, it is straightforward to compute $\eta_{unif}$ by minimizing the function $d^2(\eta) = \sum_{i=1}^{2M} (\sqrt{\eta} - \sqrt{\eta_{ii}})^2$ w.r.t. $\eta$. By computing the first derivative, the value we retrieve is nothing but the average over the diagonal elements:
\begin{equation}
    \sqrt{\eta_{unif}} = \frac{1}{2M} \sum_{i=1}^{2M} \sqrt{\eta_{ii}}
\end{equation}
Once the value for the uniform losses is provided, we may compare the results from an ideal interferometer, from an imperfect one with the loss matrix $\sqrt H$ and, eventually, from a circuit with homogeneous losses $\sqrt{\eta_{unif}} I_{2M}$. In this experiment, we adopt the Reck scheme as the optical architecture, since from the previous Section we highlighted that the Reck scheme diverges from the uniform-loss approximation. Therefore, the optical loss $\eta$ we set refers to the sole Mach-Zehnder of the optical setup, repeated across the triangular mesh of the Clements scheme, see Appendix \ref{app:Reck_loss}. 
In Figure \ref{fig:Haf}, we can see four different settings, increasing the optical losses from $\eta=0.99$ up to $\eta=0.95$, $0.90$ and $0.85$. The configuration of the parameters $\{(\theta_i, \phi_i)\}_{i=1}^{M(M-1)/2}$ is kept fixed for every $\eta$. In order to keep the value of the Hafnian constant along with $M$ in an ideal lossless setup, see Figure \ref{fig:Haf}, we set $\theta_i = \pi/4$ and $\phi_i = \pi/3$ $\forall i = 1, ..., M(M-1)/2$. 
However, the Hafnian of the matrix $A_S$ is not only determined by the angles $\{\theta_i, \phi_i \}_{i=1}^{2M}$ of the interferometer, but also by the number of photon detected in the output. For our simulation, we compute the probability to detect a configuration equal to $[2, \textbf{0}]$, with $\textbf{0} = 0^{\times M-1}$, e.g. $[2,0]$ for $M=2$, $[2,0,0]$ for $M=3$ and so on. 
Thanks to such a configuration, the value of $\mathrm{Haf(A_S)}$ remains constant, and we can see in Figure \ref{fig:Haf} how the uniform loss configuration and the realistic one diverge from each other and w.r.t. the ideal configuration. In all the reported case studies, the Hafnian computed on a device with non-uniform losses diverge more rapidly compared to the uniform case, even though both tend to converge to null values for a sufficiently high number of modes $M$, especially when decreasing the transmission rate $\eta$ for the beam-splitter in the time-bin configuration of the Reck scheme.
\begin{figure}[h]
    \centering
    \includegraphics[width=1\linewidth]{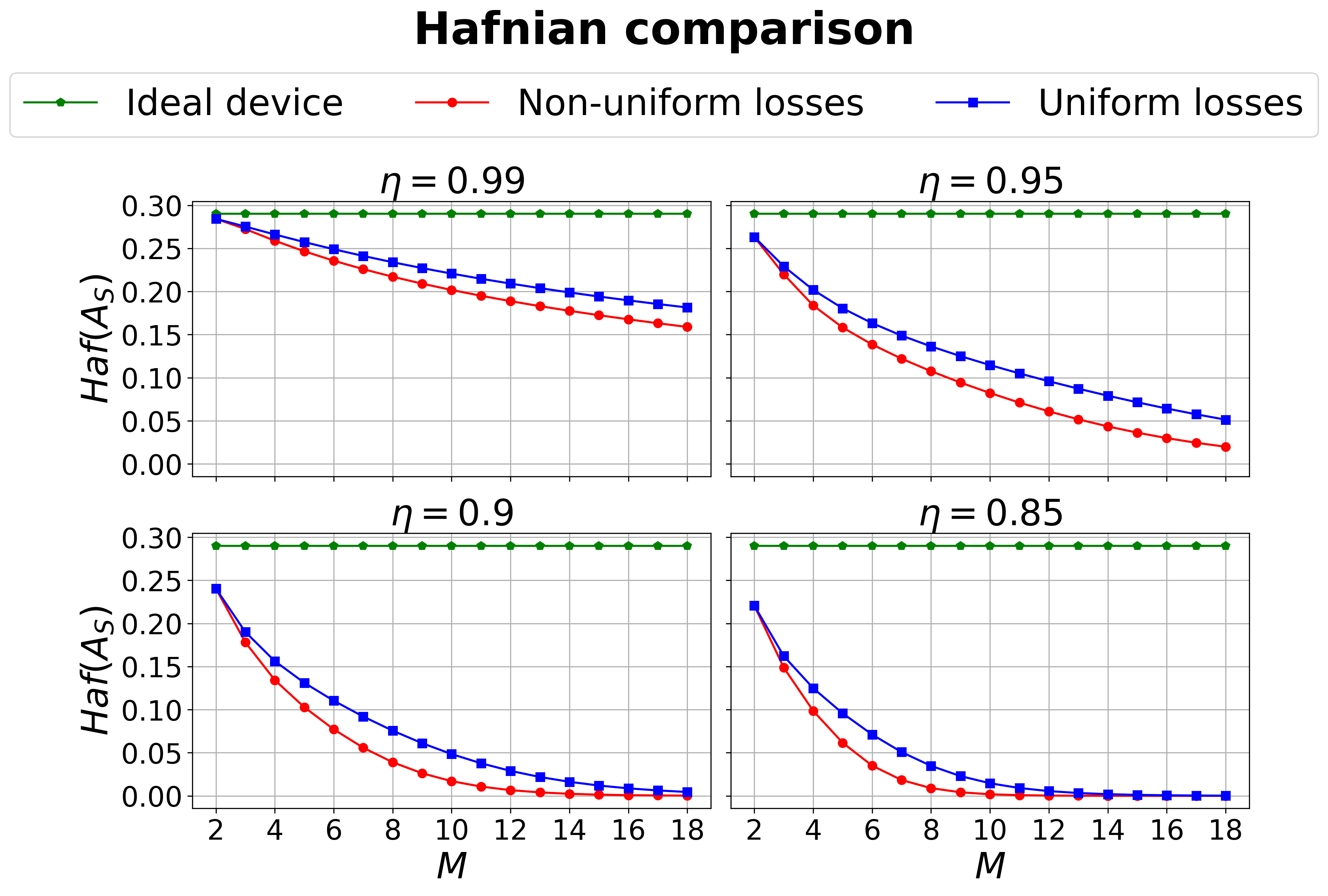}
    \caption{\label{fig:Haf}\color{black}We report the values of the Hafnian computed in an ideal device (no losses), in a device affected by uniform losses (equal $\eta$ across all the channels in the \gls{svd}) and with realistic losses, as resulting in the diagonal matrix from the \gls{svd}. The values are shown increasing the number of modes $M$. The angles of the \gls{mzi} are kept fixed to $\theta_i= \pi/4$ and $\phi_i=\pi/3$ $\forall M$ and $\forall \eta$. The value of the Hafnian is retrieved by counting the sampling configuration $[2, \textbf{0}]$.}
\end{figure}
Such a trend is highlighted in Figure \ref{fig:dis_Haf}, where the distance $d= \left| \mathrm{Haf}_{real} - \mathrm{Haf}_{hom} \right|$ between the values of the Hafnians (with uniform and realistic losses) is reported. Again, we see a net trend across all the four panels. While for $\eta=0.99$ such a trend is simply sketched, increasing the amount of optical losses it merges clearly for $\eta=0.85$. In fact, the discrepancy between the two Hafnians reaches a peak around a specific value $M_p$, after which the global effect of the optical losses consists of reducing the signal for both the Hafnians, making the two values match with zero.
\begin{figure}[h]
    \centering
    \includegraphics[width=1\linewidth]{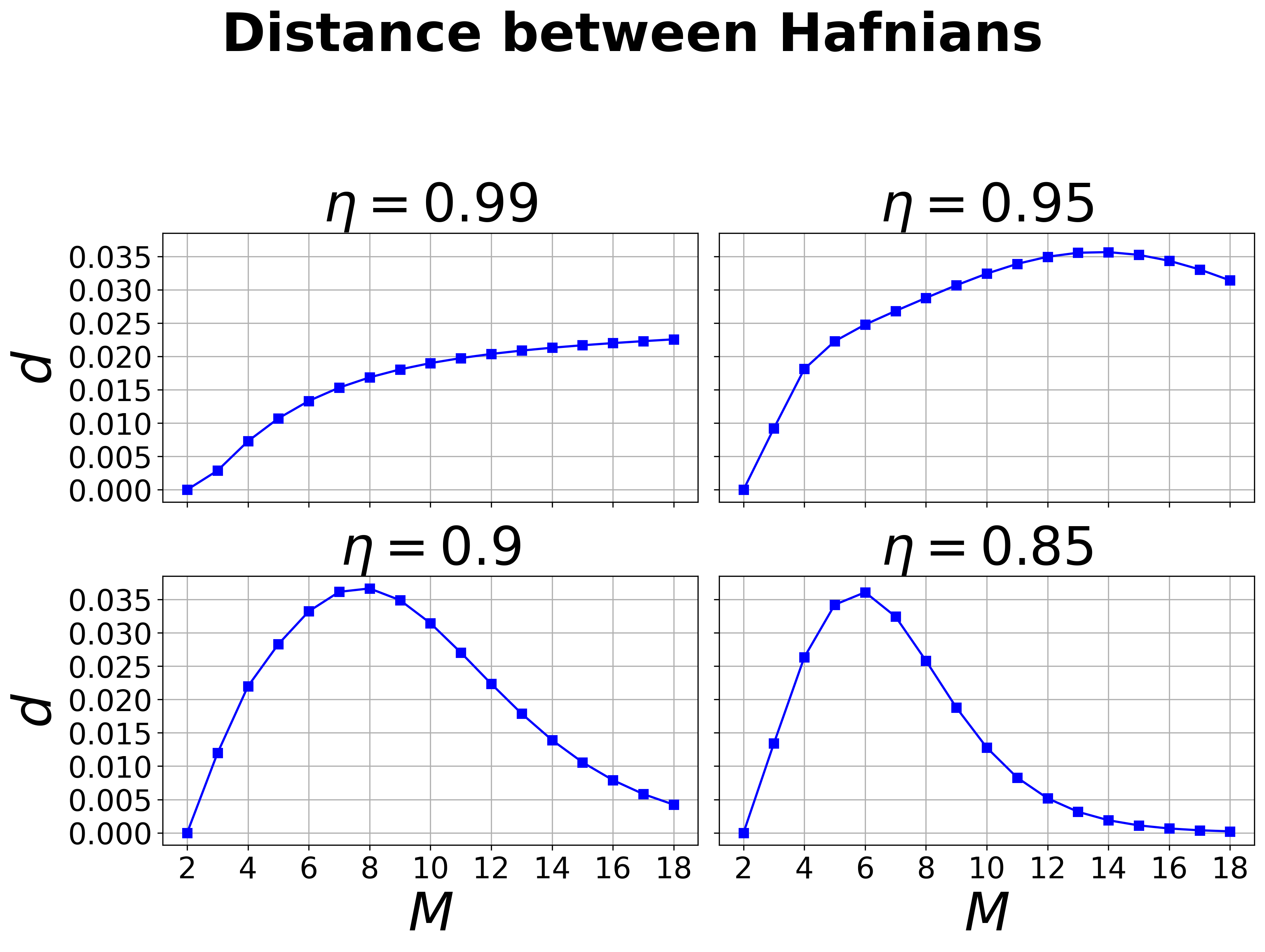}
    \caption{\label{fig:dis_Haf}\color{black}The distance $d$ (i.e. the absolute value of the difference) between the two Hafnians from the non-ideal distributions in Figure \ref{fig:Haf} is computed and reported w.r.t. with $M$ and for the four scenarios with different $\eta$.}
\end{figure}
\color{black}

\section{Conclusions}

We have proposed a framework to describe the Gaussian dynamics of a quantum optical system affected by losses, relying solely on $M$ external modes leveraging the \gls{svd}. \color{black} This main result derives from a general description of the covariance matrix of the system in presence of losses. \color{black}This formalism allows to sew empirical data on losses from the single components of the hardware on a general and theoretical frame, to describe the effects of this optical hurdle on the overall circuit.
The method is suited for both spatial and time-bin encoded architectures and for both \gls{gbs} or general-dyne-based algorithms. 
Once the dissipative factors from the single \glspl{mzi} are known, the non-unitary matrix can be numerically emulated with ease on a Gaussian backend, avoiding high-demanding simulations, both in terms of memory and operations, on the Fock space.
Nevertheless, this work aims to provide a general scheme, a blueprint to be tailor-suited for specific devices. \color{black}Within this frame, it is possible to assess with a numerical simulation whether optical losses are uniformly distributed or not, rather than assuming an \textit{a priori} distribution. \color{black}An important issue to be addressed is the characterization of optical losses due to packaging, when light enters and exits the integrated chip.
Further developments of this work may focus on simulating noise and optical losses when generating squeezed states, or even when tackling the implementation of non-Gaussian gates. As a last note, we are confident that this study could be integrated with current error mitigation techniques and that, in the field of \gls{cv} computing, it may help develop future error correction codes.

\begin{acknowledgments}
We wish to acknowledge Stefano Olivares and Alessandro Ferraro for useful discussions. EP would like to acknowledge the project QGRAPH and the project LOOQ funded by the National Centre for HPC, Big Data and Quantum Computing (D43C22001240001) for having partially supported this research. DC gratefully thanks BVTech SPA for having supported his PhD grant. 
\end{acknowledgments}

\appendix

\section{\label{app:GBS}Gaussian boson sampling}

\gls{gbs} has been introduced as an alternative version of boson sampling with single photons~\cite{hamilton2017gaussian}. \gls{gbs} techniques allow to compute the Hafnian of a sub-matrix $\mathcal{A}_s$ by sampling the number of photons from a Gaussian distribution, using squeezed states as a non-classical source. Since the evaluation of the Hafnian is known to be a $\#$P problem~\cite{bjorklund2019faster}, \gls{gbs} is envisaged among the algorithms paving the way to quantum supremacy. The general formulation of \gls{gbs} output distribution can be described as
\begin{equation}
\label{pr_gbs}
P(\bm{n}) = \frac{\mathrm{Haf}(\mathcal{A}_s)}{\bm{n}! \sqrt{\text{Det}(Q)}}
\end{equation}
with $\bm{n} = (n_1, \dots, n_M)$ accounting for the total clicks for each mode up to the $M$-th one, and $\bm{n}! = \prod_{j=1}^M n_j!$. The matrix $Q$ instead stands for the following matrix:
\begin{equation}
Q = \sigma + \frac{{I}_{2M}}{2}
\end{equation}
with ${I}_{2M}$ being the $2M \times 2M$ identity matrix, and finally $\mathcal{A}$ is defined as
\begin{equation}
\label{eq:A_def}
    \mathcal{A} = X_{2M} [{I}_{2M} - Q^{-1}]\quad ,\quad
    X_{2M} = \begin{pmatrix}
        \mathbb O & {I}_M \\
        {I}_M & \mathbb O
    \end{pmatrix}
\end{equation}
while the sub-matrix $\mathcal{A}_s$ in equation (\ref{pr_gbs}) is obtained from $\mathcal{A}$ by selecting rows and columns according to the output photon pattern $\bm{n}$. In particular, for each mode $j$, the corresponding row and column of $\mathcal{A}$ are repeated $n_j$ times. The resulting matrix $\mathcal{A}_s$ therefore has dimension $2N\times2N$, where $N=\sum_{j=1}^M n_j$ is the total number of detected photons.

When dealing with losses, $\sigma$ has to be reshaped as in Equation \eqref{eq:sigma_losses}, where the symplectic Gaussian operation matrix $\Sigma$ is that of a squeezing operation with $\{r_j\}_{j=1}^M$ as squeezing parameters~\cite{hamilton2017gaussian}, see Equation \eqref{eq:sigma_squeezing}:
\begin{equation}
\label{eq:asqueezing}
\Sigma = 
\begin{pmatrix}
    \oplus_{j=1}^M\cosh r_j &\oplus_{j=1}^M\sinh r_j\\
    \oplus_{j=1}^M\sinh r_j &\oplus_{j=1}^M\cosh r_j
\end{pmatrix}
\end{equation}
A paradigmatic example has been treated by Corollary 3, where the case of complete dissipation is discussed, to show the analytic consistency of this case with the introduced formalism.
\section{\label{app:Haar}Sampling of Haar-distributed random unitary matrices}
In this Section, we sketch how to sample a set of unitary matrices $U \in SU(M)$ according to the Haar distribution of this group. The method is properly exposed and proved in Ref. \cite{mezzadri2006generate}.
In the first place, we generate a random complex matrix $Z \in \mathbb C^{M \times M}$, with $Z_{ij} = x_{ij} + i y_{ij}$ and both $x_{ij}$, $y_{ij}$ sampled according to a Gaussian distribution $\mathcal N(0,1)$. Therefore, we perform the QR decomposition, with $Q$ belonging to the orthonormal group and $R$ being an upper triangular matrix. The next step consists of extracting the diagonal elements from $R$ and normalize them in order to achieve a set of $M$ phases, assembled in a proper $D_M$ matrix.
The final unitary matrix is nothing but $U = Q D_M$.
\section{\label{app:Reck}Decomposition of linear optical unitaries} 
In the previous Sections, we have seen that photonic computations require interferometric circuits to process quantum information. In particular, integrated photonic photonic circuits comprising networks of \glspl{mzi} have emerged as a compact and versatile solution for realizing reconfigurable linear optics. The universal decomposition of a generic linear optical unitary can be achieved using the Reck scheme \cite{reck1994experimental}, shown in Figure \ref{subfig:mesh}, a triangular mesh with each element constituting a \gls{mzi}, which is characterized by the unitary transformation in Equation (\ref{eq:interferometer}). The rectangular architecture shown in Figure \ref{subfig:mesh} is also universal, and benefits from improved compactness \cite{Bell_2021}. These two decomposition architectures are crucial in simulating a real photonic computation, since we can decompose the unitary matrix we want to implement via one of these two schemes, and then add the photon loss phenomena on the system, multiplying each \gls{mzi} matrix with its respective loss matrix, as in Equation (\ref{eq:loss_interf_big}).
\subsection{Overview on the two architectures}
In the Reck and Clements scheme, each lower off-diagonal entry of the target unitary matrix is sequentially nullified using a $2\times2$ \gls{mzi} matrix, as the operators in Equation \eqref{eq:loss_interf_big} (with $\eta=1$, since we do not care about losses in the decomposition). Since a $M$-dimensional matrix consists of $M(M-1)/2$ lower off-diagonal elements, we require the same number of interferometers.
Upon nullifying all lower off-diagonal entries, at the end of the process we are left with an upper triangular matrix, but due to the constraints of the unitary group, such matrix reduces to a diagonal one denoted by
\begin{equation}
D_M(\alpha_1,\dots, \alpha_M) =
\begin{pmatrix}
    \exp[i\alpha_1] & &\\
     &\ddots &\\
     & & \exp[i\alpha_M]
\end{pmatrix}
\end{equation}
This diagonal matrix can be realized by applying $M$ phase-shifters \cite{yasir}.
In the Reck scheme, the last row of the target unitary (except the $U_{M,M}$ entry) is first nullified. Thus, the next step is to nullify the $(M-1)$-th row up to the $U_{M-1, M-1}$ entry. Proceeding further, this nullification process gives rise to a triangular architecture, as we see in Figure \ref{subfig:mesh}. A practical example of decomposition for a $3 \times 3$ matrix is illustrated in Equation \eqref{eq:reck_3}.
On the other hand, the nullification process in the Clements scheme occurs in the following order of entries: $U_{M,1}$\,, $U_{M-1,1}$\,, $U_{M,2}$\,, $U_{M ,3}$\,, $U_{M-1 ,2}$\,, and so on. This process of nullification leads to a rectangular architecture~\cite{Clements, yasir}, as we see again in Figure \ref{subfig:mesh}.
Therefore, any $M$-dimensional unitary matrix $U_M$ can be decomposed as
\begin{equation}
U_M = D_M\prod_{m,n}T_{mn}(\theta_{mn},\phi_{mn})
\label{decomposition}
\end{equation}
where $T_{mn}$ is a $M$-dimensional identity matrix with entries at positions $(m,m)\,,(m,n)\,,(n,m)\,,(n,n)$ be replaced by the entries of the \gls{mzi} matrix in Equation (\ref{eq:interferometer}).
\subsection{Pipeline on decomposing via the Reck scheme}
In order to decompose a $M$-dimensional unitary matrix, we have to find the set of parameters $\{(\theta_j, \phi_j)\}_{j=1}^{M(M-1)/2}$ and $\{\alpha_j\}_{j=1}^M$ which satisfy Equation (\ref{decomposition}), $\alpha_j$ being the phases of the diagonal matrix $D_M$.
To do so, first of all we care about the parameters $\theta$ and $\phi$ which define the $M(M-1)/2$ interferometers: via the Reck scheme the inverse matrices of such interferometers acts on the target unitary matrix on the right \cite{reck1994experimental}, as we see in Equation (\ref{eq:reck_3}), in which we we show an example of nullifying operation of a $3\times3$ unitary matrix:
\begin{equation}
\label{eq:reck_3}
    \begin{pmatrix}
        u_{11} & * & * \\
        \color{olive} u_{21} & u_{22} & * \\
        \color{red}u_{31} & \color{blue} u_{32} & u_{33}
    \end{pmatrix}
    \,\color{red}T_{1,2}^\dagger(\theta_1,\phi_1) \color{blue} T_{2,3}^\dagger(\theta_2,\phi_2) \color{olive} T_{1,2}^\dagger(\theta_3,\phi_3) \color{black} = D_3
\end{equation}
Thus every right-nulling operation can be reduced to the product of two $2\times2$ square matrices. For the first operation e.g.:
\begin{equation}
\begin{pmatrix}
    u_{11} &u_{12} & u_{13}\\
    u_{21} &u_{22} & u_{33} \\
    u_{31} &u_{32} & u_{33} \\
\end{pmatrix}
\begin{pmatrix}
    e^{-i\phi}\cos\theta &e^{-i\phi}\sin\theta & 0\\
    -\sin\theta &\cos\theta & 0 \\
    0 & 0 & 1
\end{pmatrix} = 
\begin{pmatrix}
    u'_{11} & u'_{12} & u'_{13}\\
    u'_{21} & u'_{22} & u'_{33} \\
    0 & u'_{32} & u'_{33} \\
\end{pmatrix}
\end{equation}
such that the element of the unitary matrix we want to nullify becomes
\begin{equation}
u_{31}\mapsto u_{31}e^{-i\phi}\cos\theta - u_{32}\sin\theta
\end{equation}
Parametrizing the unitary matrix elements $a=u_{31}$ and $b=u_{32}$, in order to evaluate the \gls{mzi} angles which perform the nullification, we have to solve the equation
\begin{equation}
ae^{-i\phi}\cos\theta = b\sin\theta
\end{equation}
such that we find
\begin{subequations}
\label{eq:right_nulling}
\begin{align}
&\theta = \arctan\Big(\frac{|a|}{|b|}\Big)\\
&\phi = \arg(a) - \arg(b)
\end{align}
\end{subequations}
A proper Python function \texttt{right\_nulling} must take the parameters $a,b$ and evaluate the two angles $\theta,\phi$ following Equation \eqref{eq:right_nulling}. Thus, the pipeline to evaluate the set of parameters $\bm{\alpha}=\{\alpha_j\}_{j=1}^M$ and $(\bm \theta, \bm \phi) =\{(\theta_j, \phi_j)\}_{j=1}^{M(M-1)/2}$ of the Reck decomposition is shown in Algorithm \ref{alg:reck}.\\[0.4cm]
\begin{algorithm}[H]
\SetAlgoLined
\KwIn{$U \in \mathbb{C}^{M \times M}$}
\KwOut{$\bm{\alpha}$, $(\bm \theta, \bm \phi)$}
$counter = M-1$\;
$idx = 0$\;
\While{$counter \geq 1$}{
    \For{$mode = 0$ \KwTo $counter-1$}{
        $a \leftarrow U_{counter, mode}$
        
        $b \leftarrow U_{counter, mode+1}$

        $(\theta,\phi) = \texttt{right\_nulling}(a, b)$

        $U \leftarrow U \, T^\dagger_{mode}(\theta,\phi)$

        $(\theta_{idx},\phi_{idx}) \leftarrow (\theta, \phi)$ 

        $idx \leftarrow idx + 1$
    }
    $counter \leftarrow counter - 1$
}
$D \leftarrow U$
$\alpha_j = \arg(D_{jj})$
\Return{$\boldsymbol{\alpha}$, $(\bm \theta, \bm \phi)$}
\caption{\label{alg:reck}Reck decomposition of a unitary matrix}
\end{algorithm}

\section{\label{app:Reck_loss}Optical losses in the loop-based Reck scheme}

\begin{figure}
    \centering
    \subfloat[]{\includegraphics[width=0.35\linewidth]{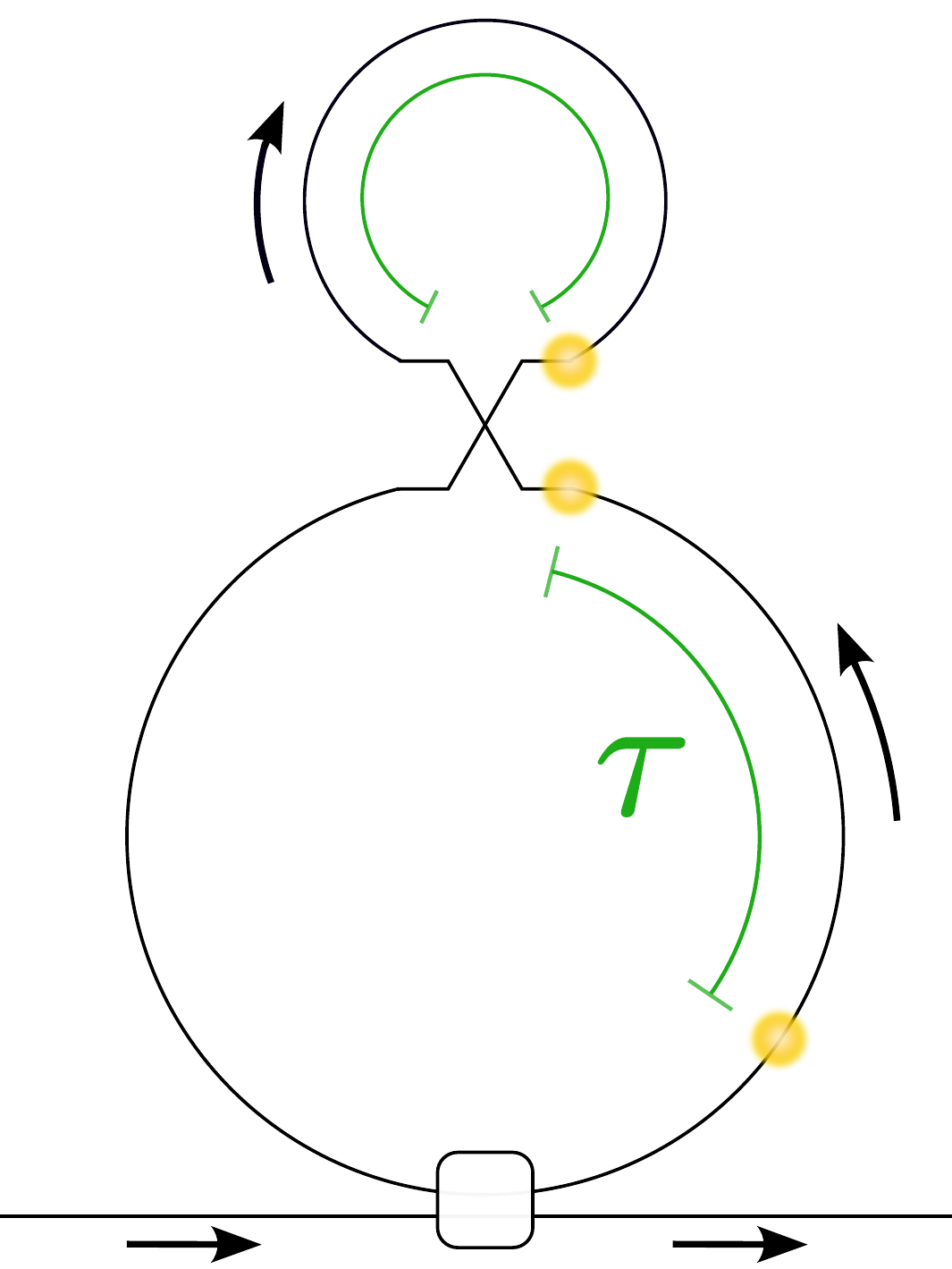}\label{subfig:Mozzarella}}
    \hfil
    \subfloat[]
    {\includegraphics[width=0.35\textwidth]{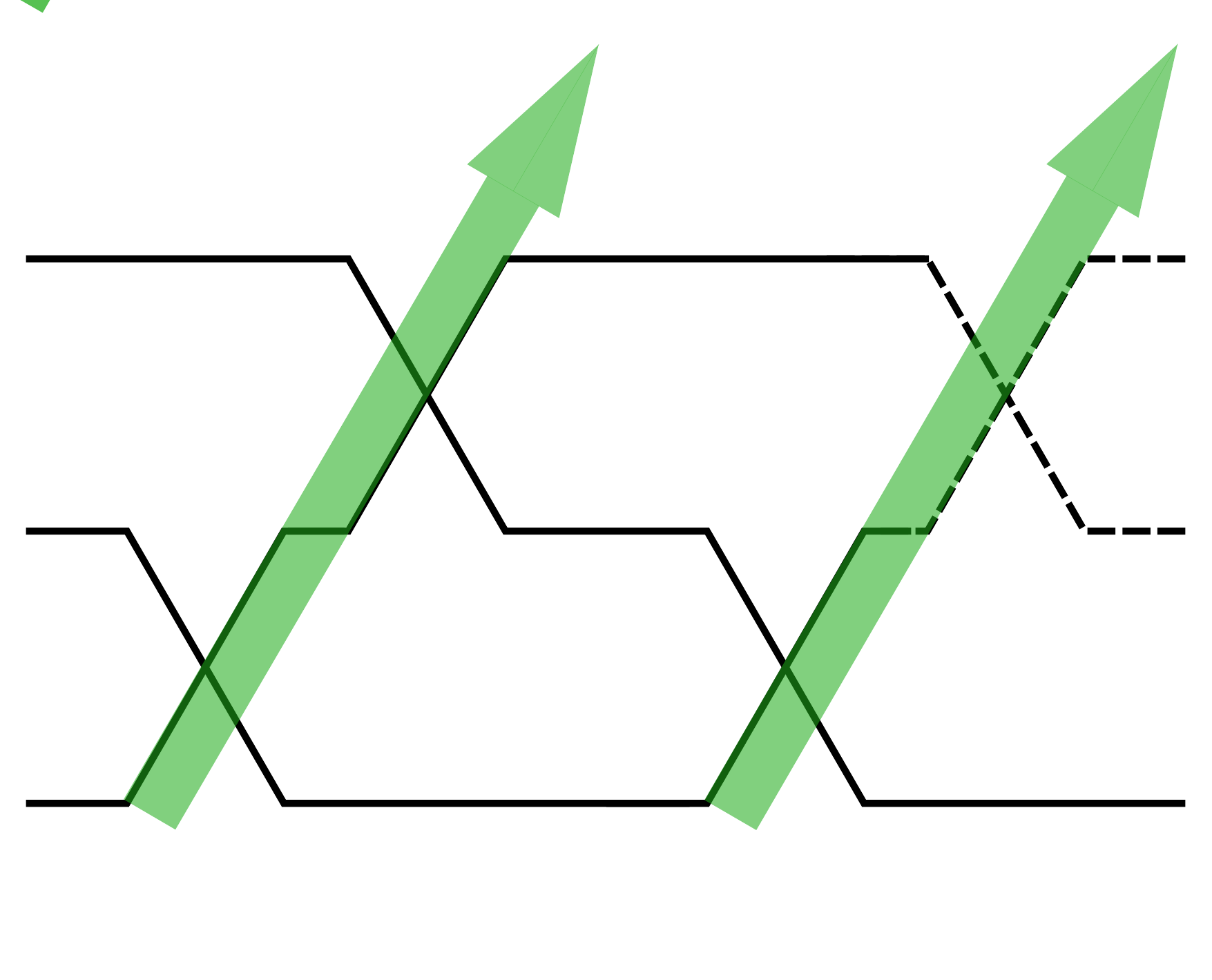} \label{subfig:Reck_loop}}
    \caption{\label{fig:Reck_loop_losses} (a) The loop architecture and (b) its equivalent mesh transposition. In an ideal system with no losses, the equivalent spatial mesh matches with the Reck scheme. In the loop architecture, the wave packets are introduced in the big loop by a switch (rectangle in the bottom). The temporal distance $\tau$ between them equals the time they circle into the small loop. Their interaction happens through an interferometer when respectively entering and exiting the small loop. The green arrows correspond to a loop cycle, i.e. the time lapse from the first packet entering the small loop till the last one exiting it. The dashed interferometers correspond to operations equal to the identity, except for the losses induced by the passage of the packets through the loop.}
\end{figure}

In this Section, we illustrate how optical losses affect the loop-based computation, starting from the Reck scheme. In fact, since the decomposition engineered by Clement provides a more compact mesh for spatial encoding, the Reck scheme nowadays mainly refers to time-bin encoding approaches.
We refer to Figure \ref{subfig:Mozzarella} to better explain this frame. First, the wave packets enter the big loop through a switch. All packets are separated equally by a time interval $\tau$, which corresponds to the time lapse for a packet to run inside the small loop.
The big and small loops are interconnected by a \gls{mzi}. If the interferometer acts as the identity, e.g. with $\theta=\phi=0$ following the convention adopted in Equation \eqref{eq:interferometer}, the packets inside the small loop are confined inside of it, as for the packets inside the big one. Otherwise, once the first packet enters the small loop, it interacts after $\tau$ with the second packet coming from the big loop, which interacts afterward with the third one and so on until the $(n-1)$-th interacts with the $n$-th one.
In this scheme, the wave packets correspond to the computational modes.
Once all packets have entered the small loop, the first diagonal of \glspl{mzi} in the Reck scheme has been implemented. Of course, the interferometer connecting the two loops is required to be reprogrammable in a very short time. Nowadays, lithium niobate modulators rely on the Pockels effect to tune the \gls{mzi} angles, which requires around few femtoseconds~\cite{zhao2025sub}.
In Figure \ref{subfig:Reck_loop}, the green arrow represents the implementation of the first $n-1$ \glspl{mzi} in the first loop cycle. Once the first wave packet has circled inside the big loop and entered the small loop for the second time, we require it to interact with the second mode and so on until the $(n-2)$-th mode interacts with the $(n-1)$-th, leaving unaltered the $n$-th. The second diagonal in the Reck scheme, now made up of $n-2$ interferometers, has been implemented. In Figure \ref{subfig:Reck_loop}, the second loop cycle corresponds to the second arrow. However, since the $n$-th packet has to be switched with the $(n-1)$-th one to allow it to leave the small loop, an additional interferometric operation is required, even though computationally negligible. Nevertheless, this last operation involves additional losses, reason why an additional dashed interferometer appears in Figure \ref{subfig:Reck_loop}.

\section{Importance of the environmental modes}

In this Appendix, we show why external modes are mandatory for a correct formulation of the covariance matrix.
When introducing non-idealities, we can perform the \gls{svd} and resume Equation \eqref{eq:SVD_losses} to describe the dynamics of the quantum system:
\begin{equation}
\label{eq:decomp}
    S = \mathcal U \Sigma = \mathcal U_2 \sqrt H \mathcal U_1 \Sigma
\end{equation}
If neglecting the external modes, such an approach would lead to a peculiar formulation of  the covariance matrix:
\begin{equation}
\label{eq:wrong_cov}
    \sigma = \frac{1}{2} S S^\dagger \Rightarrow
    \sigma = \frac{1}{2} \mathcal U_2 \sqrt H \mathcal U_1  \Sigma^2 \mathcal U_1^\dagger \sqrt H \mathcal U_2^\dagger 
\end{equation}
We now handle two use cases we have introduced in the Corollaries from Section \ref{sec:cov_mat}, to highlight the incorrectness of the expression in Equation \eqref{eq:wrong_cov}. In case of total dissipation, the covariance matrix in fact would read as $\sigma = \mathbb O$, a value which has no physical meaning, since this matrix should equal the covariance matrix of the vacuum state.
The second case showcases an imperfect interferometer with no squeezed light, i.e. another vacuum state. The expression from Equation \eqref{eq:wrong_cov} would lead to $\sigma = \mathcal U_2 H \mathcal U_2^\dagger$, which again differs from the covariance matrix of the vacuum state.



\bibliography{apssamp}

\end{document}